\documentclass[12pt, draftclsnofoot, onecolumn]{IEEEtran}

\usepackage{booktabs}
\usepackage[square, comma, sort&compress, numbers]{natbib}
\usepackage{bbm}
\usepackage{amssymb,amsmath,amsfonts}
\makeatletter
\def\maketag@@@#1{\hbox{\m@th\normalfont\normalsize#1}}
\makeatother
\usepackage{graphicx}
\usepackage{textcomp}
\usepackage{xcolor}
\usepackage{diagbox}
\usepackage[misc]{ifsym}
\usepackage{mathrsfs}
\usepackage{float}
\usepackage{multirow}
\usepackage{subfigure}
\usepackage{titlesec}
\usepackage{url}
\usepackage{bm}
\usepackage[numbers]{natbib}
\usepackage{algpseudocode}
\usepackage{epstopdf}
\usepackage{makecell}
\usepackage{color}
\usepackage[ruled,vlined]{algorithm2e}
\floatname{algorithm}{Algorithm}

\allowdisplaybreaks[4]

\titlespacing{\section}{0pt}{0ex}{0pt}
\titlespacing{\subsection}{0pt}{0ex}{0ex}
\titlespacing{\subsubsection}{0pt}{0ex}{0pt}

\begin{document}

\title{Joint Vehicular Localization and Reflective Mapping Based on Team Channel-SLAM}

\author{
        Xinghe~Chu,
        Zhaoming~Lu*,
        David~Gesbert,~\IEEEmembership{Fellow,~IEEE},
        Luhan~Wang,
        Xiangming~Wen,
        Muqing Wu,
        Meiling Li
%
%         % <-this % stops a space
%
\thanks{

Xinghe Chu, Zhaoming Lu, Luhan Wang, Xiangming Wen and Muqing Wu are with Beijing Laboratory of Advanced Information Networks, Beijing University of Posts and Telecommunications, Beijing, China, e-mails: \{chuxinghe, lzy0372, wluhan, xiangmw, wumuqing\}@bupt.edu.cn ({\em{Corresponding author: Zhaoming Lu}}).
David~Gesbert is with the Communications Systems Department, EURECOM, Sophia Antipolis, France, email: David.Gesbert@eurecom.fr.
Meiling Li is with Taiyuan University of Science and Technology, Shanxi, China, email: meilingli@tyust.edu.cn.
This paper is supported by Beijing Nova Program under grant Z201100006820123. The work of D. Gesbert was partially funded via the HUAWEI France supported Chair on Future Wireless Networks at EURECOM.}% <-this % stops a space

}

\markboth{Joint Vehicular Localization and Reflective Mapping Based on Team Channel-SLAM}%
{Shell \MakeLowercase{\textit{et al.}}: Bare Demo of IEEEtran.cls for IEEE Journals}

% make the title area
\maketitle

\vspace{-3ex}
\begin{abstract}

This paper addresses high-resolution vehicle positioning and tracking.
In recent work, it was shown that a fleet of independent but neighboring vehicles can cooperate for the task of localization by capitalizing on the existence of common surrounding reflectors, using the concept of Team Channel-SLAM. This approach exploits an initial (e.g. GPS-based) vehicle position information and allows subsequent tracking of vehicles by exploiting the shared nature of virtual transmitters associated to the reflecting surfaces. In this paper, we show that the localization can be greatly enhanced by joint sensing and mapping of reflecting surfaces. To this end, we propose a combined approach coined Team Channel-SLAM Evolution (TCSE) which exploits the intertwined relation between (i) the position of  virtual transmitters, (ii) the shape of reflecting surfaces, and (iii) the paths described by the radio propagation rays, in order to achieve high-resolution vehicle localization.
Overall, TCSE yields a complete picture of the trajectories followed by dominant paths together with a mapping of reflecting surfaces.  While joint localization and mapping is a well researched topic within robotics using inputs such as radar and vision, this paper is first to demonstrate such an approach within mobile networking framework based on radio data.

\end{abstract}


\IEEEpeerreviewmaketitle

\section{Introduction}

% \IEEEPARstart{key}\\
% \\

\IEEEPARstart{T}{he} fifth-generation wireless networks aim to support demanding services such as enhanced Mobile Broadband, Ultra-Reliable and Low Latency Communications, as well as massive Machine-Type Communications with enhanced data rate and few milliseconds' latency \cite{navarro2020survey,li20185g,agiwal2016next,gupta2015survey,niu2015survey}.
Many new services will also require high accuracy localization capabilities, for instance in the domain of network-assisted intelligent transport, autonomous vehicles and robots.
  More recently the use of radio signals provided by the mobile network itself have been investigated as means to enable the localization of user equipments \cite{wen2019survey}. In the context of 5G networks, the resolution capabilities of such methods have improved dramatically
thanks to the use of large bandwidth and massive Multiple Input and Multiple Output system \cite{witrisal2016high,menta2019performance,keating2019overview,3gpp_TS_38_211}. The works \cite{horiba2015accurate,mingyang2008distributed,jiao2009lcrt,Marano2010NLOS,del2017survey} that rely on the estimation of Time of Arrival (ToA), Angle of Arrival (AoA) and Angle of Departure (AoD) have provided possible strategies for radio localization. However, much of this related work tends to specialize to scenarios where the Line of Sight (LoS) path is dominant. In practice, it is well known that the sensibility to (and ignorance of the information carried by) multi-path components limits the localization accuracy \cite{witrisal2016high}.
To cope with this problem, the multi-path assisted positioning methods such as \cite{wang2011omnidirectional,han2019hidden} offer an interesting alternative and they allow for the exploitation of the spatial information carried by multi-path components for localization.
In principle, 5G radio-based localization methods that can leverage multi-path offer substantial advantages over classical approaches (GPS, RADAR and LiDAR, etc.) as they are naturally more robust with respect to an obstructed propagation environment. They are also cost-effective as they reuse existing components and devices without the need for extra hardware \cite{ko2021v2x}.
While methods like \cite{wang2011omnidirectional,han2019hidden} capitalize on the multi-paths to localize the user, they cannot exploit the underlying structure that link the multi-path to the environment.

To this end, the Channel Simultaneous Localization and Mapping (Channel-SLAM)  \cite{gentner2016multipath,wymeersch20185g,mendrzik2018joint,palacios2017jade,palacios2018communication,aladsani2019leveraging,yassin2018mosaic,mendrzik2019enabling,leitinger2019belief,gentner2021simultaneous,meng2021v2v} methods were proposed for radio localization, which can exploit the link between the radio paths and the static reflecting surfaces which create them. Channel-SLAM methods work by allowing to exploit the static nature of the environment by recasting the multi-path components as pseudo LoS paths that originate from a set of quasi-static virtual transmitters (VT) to improve the localization accuracy.  Hence by estimating the location of those VTs it is possible to improve the user localization.  As a natural extension of the initial Channel-SLAM concept \cite{gentner2016multipath}, 
{{the cooperative radio-SLAM methods referred to here as Team Channel-SLAM \cite{chu2020team,chu2021vehicle,kim20205g}  exploit}}
the multiple user nature of road traffic to provide a cooperative approach to positioning vehicles in a NLoS environment.

More precisely, Team Channel-SLAM methods leverage the fact that neighboring vehicles will be surrounding by the same reflecting surfaces. The algorithm can then explore the co-dependence between the virtual transmitters observed by these vehicles by building a  Common Virtual Transmitter (CVT) model. This way it is possible to improve the localization accuracy over a single vehicle scenario, with performance growing with the vehicle density.
The topic of multi-vehicle localization is rich with contributions from the existing literature. In some approaches, the concept of multiple-target tracking based on radio measurements and VTs is exploited such as in \cite{kim20205g,meyer2018message,soldi2019self} which shows some similarities with our work. However substantial differences remain. The most striking one is that in previous methods that use radio data to construct information about VTs, the information related to the surfaces that cause the reflection is kept implicit, hence not fully leveraged for the vehicular localization purpose.
It should be also noted that such existing methods typically need an initial (e.g. GPS-based) position input {{in a single base station case}} for each new-coming vehicle in order to
activate the algorithm, which can be hard to obtain when the satellite signal is obstructed by high-rise
buildings or bridges. One more advantage of reconstructing surface information as part of the algorithm in our case is that this initial position estimate derives naturally from the algorithm.

In this paper, we propose a more comprehensive approach
coined Team Channel-SLAM Evolution (TCSE). TCSE is a closed-loop approach allowing to exploit the inter-relations between information residing the surrounding reflective structures, the shape of reflecting surfaces and the radio path geometry (so-called radio geometrization) for vehicular localization, as illustrated in Fig. \ref{fig_VT}.
Our framework also includes wake-up positioning and synchronization\footnotemark[1] aided by radio geometrization, which provides an accurate initial position and time synchronization input for cooperative SLAM. 
{{Though the works \cite{wymeersch20185g,9179819,8517150,8515231} also consider synchronization under Channel-SLAM framework, the information of reflecting surfaces is not considered together in their synchronization bias estimation process.}}
This also makes it more robust to satellite-signal conditions that typically impede accurate GPS-based localization in real-life situations, such as the presence of high-rise buildings or other obstructions. Besides vehicle positioning, the proposed framework can also be exploited for the benefit of improving communication performance. For instance, the information about dominant paths extracted from radio geometrization can be used to optimize beamforming and beam-alignment strategies \cite{zeng2021toward,koivisto2017high,di2014location,aviles2016position,sachan2016genetic}. The proposed TCSE approach encompasses three main components which are presented below:

\footnotetext[1]{Wake-up positioning and synchronization refers to estimating the position and time synchronization bias of a newcomer target (target just entering the communication range of a base station without its position and time synchronization known).}

{\em Cooperative SLAM:}
Cooperative SLAM explores the shared nature of VTs among multiple neighboring vehicles through a CVT model.
For tracking purposes, the CVT model are maintained by a belief propagation (BP) based data association algorithm. BP was previously proposed in the context of user localization {{\cite{leitinger2019belief,meyer2018message,mendrzik2019enabling,leitinger2020data,kim20205g,9443465}}}, however not combined with the estimation of reflective surfaces to which CVTs are associated as in our paper. Also our approach is designed to detect and handle false alarms (FA) in the association process based on estimated reflecting surfaces, 
{{while such information from reflecting surfaces is not considered in \cite{leitinger2019belief,meyer2018message,mendrzik2019enabling,leitinger2020data,kim20205g,9443465}.
}}
The probabilistic distributions for the positions of vehicles and CVTs are estimated in the form of discrete particles by a team particle filter.

{\em Reflective Sensing and Mapping:}
Reflective sensing and mapping extracts the information of reflecting surfaces from the cooperative SLAM. More precisely, it estimates the position and edge of reflecting surfaces, which are provided for the other two components. 
{{Though \cite{leitinger2020data} also considers wall features rather than directly VTs, it fails to explore their relation with radio geometrization and FA handling.}}
For this, we use the classical  online learning approach called Follow the Regularized Leader (FTRL) \cite{duchi2011adaptive}.

{\em Radio Geometrization:}
Radio geometrization estimates the paths described by the radio propagation rays through the Viterbi \cite{forney1973viterbi} based reflector decoding algorithm and achieves wake-up positioning and synchronization\footnotemark[1].
This component can provide precise initial position and time synchronization for the cooperative SLAM component 
{{using only one base station, which does not need the synchronization between multiple base stations}}.

The main purpose of this paper is to present a comprehensive method for vehicle positioning and tracking referred to as TCSE, which utilizes the radio signal within a mobile networking framework to simultaneously build a 3-D map of reflecting surfaces to improve the accuracy of vehicular localization and its robustness to satellite-positioning signal conditions.
The contributions of this paper can be summarized as follows:

\begin{itemize}
  \item  We introduce the Team Channel-SLAM Evolution method to achieve simultaneous multiple vehicle localization and CVT positioning in an inter-vehicle cooperative manner together with reflective sensing and mapping as well as radio geometrization.

  \item  Our reflective sensing and mapping method allows one to estimate the 3-D position and edge of the reflecting surfaces based on the reflecting elements extracted from the SLAM procedure.

  \item  In contrast with previous SLAM-based positioning methods, our radio geometrization method allows to estimate the geometry of all dominant paths between the base station and the vehicles while bypassing the need for initial position and time synchronization.

\end{itemize}

The remaining sections are organized as follows.
Section \ref{sec_system_model} introduces the system models. Section \ref{sec_CO_SLAM} introduces the cooperative SLAM component. Section \ref{sec_environmental_awareness} introduces the reflective sensing and mapping component. Section \ref{sec_radio_geometrization} introduces the radio geometrization component. Section \ref{sec_implementation} introduces the overview of  implementation for TCSE.
Simulations are done in Section \ref{sec_numberical_results} and the conclusion is drawn in Section \ref{sec_conclusion}.

{\em Notations.} Vectors and matrixes are displayed in bold type. Operator $ \otimes $ denotes the Kronecker product, $\left\|  \cdot  \right\|_F$ denotes the Frobenius norm,  $\left|  {\bm a}  \right|$ denotes the number of elements in vector ${\bm a}$, ${\left(  \cdot  \right)^T}$ denotes the transposition for a matrix,  ${\left(  \cdot  \right)^{ - 1}}$ denotes the inversion of a matrix, $\left\lfloor  \cdot  \right\rfloor $ denotes the nearest integer less than or equal to that element,  $\oplus$ denotes the xor operation,
${\mathbb E}\left(  \cdot  \right)$ denotes the expectation of a variable and $diag\left(  {\bm x}  \right)$ denotes the diagonal matrix with the elements of ${\bm x}$ on its diagonal and zero else-where. The variables with upper arc-shaped ${\overset{\lower0.5em\hbox{$\smash{\scriptscriptstyle\frown}$}}{ \cdot } }$ denote the estimation of corresponding variables, and $p\left(  \cdot  \right)$ denotes the probability density function.
  We write $k$ for the index of discrete time index, $m$ for the index of vehicles, $p_m$ for the index of multi-path components observed by vehicle $m$, $n$ for the index of CVTs, $l$ for the index of reflectors, $i$ for the index of vehicle particles, and $j$ for the index of CVT particles. Specially, we use $\left\{  \cdot  \right\}$ to denote the set of corresponding indexes for $m,p_m,n,l$, e.g. $\left\{ l \right\} \triangleq  \left\{ {l\left| {l = 1,2,...L} \right.} \right\}$ means the set of all the reflector indexes.

\section{System Model}
\label{sec_system_model}
\subsection{Vehicle State Model}
\begin{figure}[!t]
  \setlength{\abovecaptionskip}{-0.15cm}
  \centering
  \includegraphics[width=0.45\columnwidth]{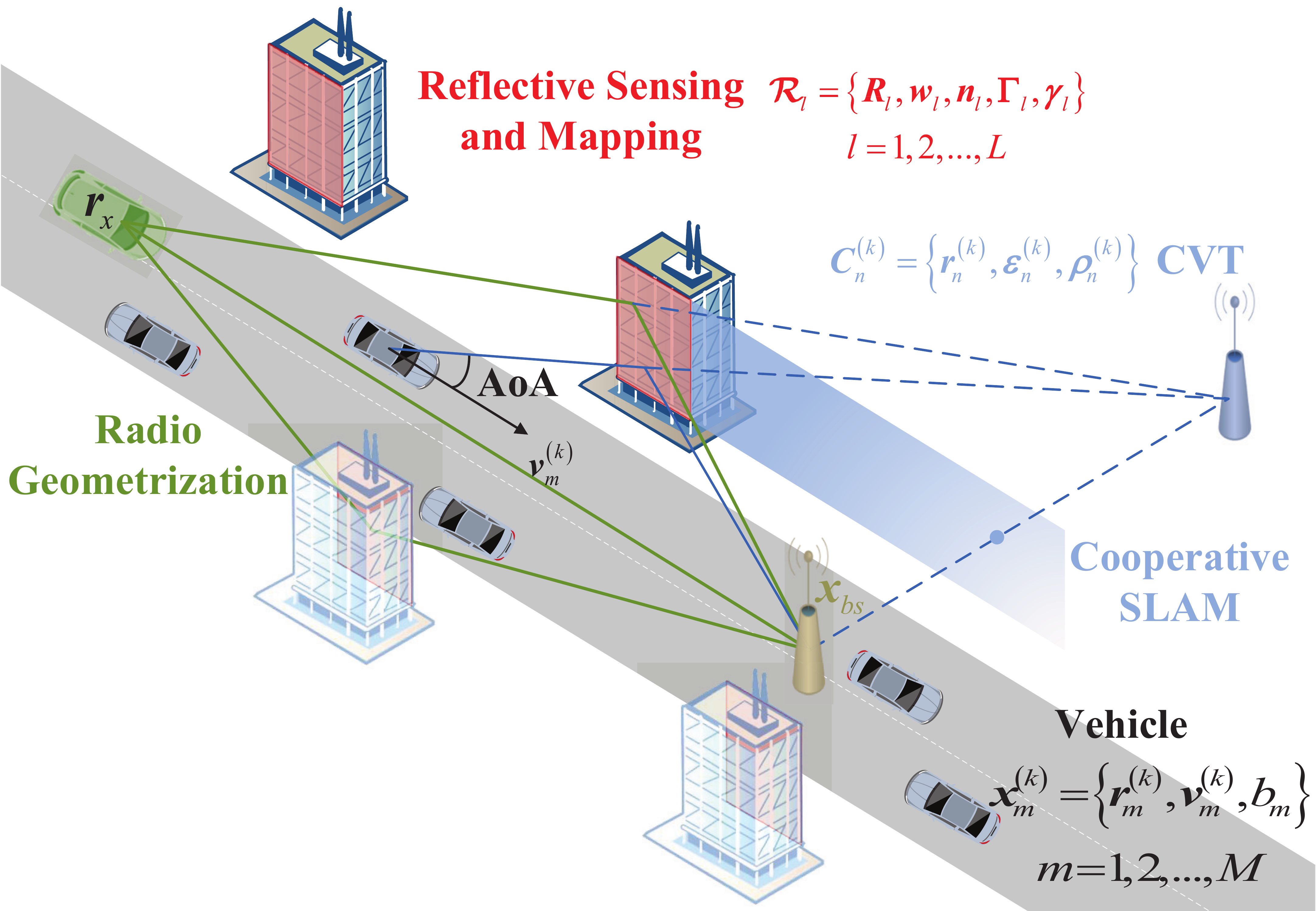}
  \caption{System models for cooperative SLAM,  reflective sensing and mapping, and radio geometrization in TCSE.
  }
  \label{fig_VT}
  \vspace{-5ex}
\end{figure}

As shown in Fig. \ref{fig_VT}, the location of the base station is denoted as ${\bm x}_{bs}$, and there are $M$ vehicles, where the state for the $m$-th vehicle at time $t_k$ can be denoted as ${\bm x}_m^{\left( k \right)} = \left\{ {{\bm r}_m^{\left( k \right)},{\bm v}_m^{\left( k \right)},b_m} \right\}$.
${\bm r}_m^{\left( k \right)}$ and ${\bm v}_m^{\left( k \right)}$ denote the position and velocity for vehicle $m$, respectively.
 ${ b}_m$ denotes the time synchronization bias for vehicle $m$, which is assumed to be constant but unknown for each vehicle. Specially, the probabilistic distribution of ${\bm r}_m^{\left( k \right)}$ is approximate by  discrete particles denoted as $p\left( {{\bm r}_m^{\left( k \right)},{b_m}} \right) \triangleq \left\{ {p\left( {{\bm r}_m^{\left( k \right)} = {\bm r}_m^{\left( {k,i} \right)},{b_m} = b_m^{\left( {k,i} \right)}} \right) = w_m^{\left( {k,i} \right)}} \right\}_{i = 1}^{{{\cal N}_V}}$,
where ${w_m^{\left( {k,i} \right)}}$ is the weight of the $i$-th particle, ${{\bm r}_m^{\left( {k,i} \right)}}$ and ${b_m^{\left( {k,i} \right)}}$ denote the position and time synchronization bias of the $i$-th particle, and ${{{\cal N}_V}}$ denotes the number of vehicle particles.
\subsection{Observation Model and Virtual Transmitter Model}
\label{sec_OMVTM}
The observations from vehicle $m$ at time $t_k$ include ToA and AoA measurements for each multi-path component of the signals from the  base station to that vehicle {{as the case in \cite{8756824}}}, which can be denoted as (ToA is multiplied by the speed of light, and AoA is shown in Fig. \ref{fig_VT}):
\begin{equation}\setlength{\abovedisplayskip}{3pt}\setlength{\belowdisplayskip}{3pt}
  \begin{array}{c}
{\bm z}_m^{\left( k \right)} = \left\{ {{\bm z}_{\left( {m,1} \right)}^{\left( k \right)},...,{\bm z}_{\left( {m,{p_m}} \right)}^{\left( k \right)},...,{\bm z}_{\left( {m,{P_m}} \right)}^{\left( k \right)}} \right\},\
{\bm z}_{\left( {m,{p_m}} \right)}^{\left( k \right)} = \left\{ {d_{\left( {m,{p_m}} \right)}^{\left( k \right)},\theta _{\left( {m,{p_m}} \right)}^{\left( k \right)},\varphi _{\left( {m,{p_m}} \right)}^{\left( k \right)}} \right\},
\end{array}\label{eq_SM_4}
\end{equation}
where ${\bm z}_m^{\left( k \right)}$ is the observation from vehicle $m$ at time $t_k$, ${\bm z}_{\left( {m,{p_m}} \right)}^{\left( k \right)}$ is the observation of its $p_m$-th multi-path component, and ${d_{\left( {m,{p_m}} \right)}^{\left( k \right)},\theta _{\left( {m,{p_m}} \right)}^{\left( k \right)},\varphi _{\left( {m,{p_m}} \right)}^{\left( k \right)}}$ denote the measurements of ToA, azimuth angle of AoA, and polar angle of AoA, respectively.
We assume that the ToA and AoA have been previously estimated with errors following Gaussian distributions \cite{mendrzik2019enabling,leitinger2019belief}. The AoA estimations are obtained by a uniform planner array on the vehicle \cite{fazliu2020graph} with zero mean error, and the mean error of ToA estimation is constant and determined by the time synchronization bias $b_m$.

Each multi-path component can be recast into a LoS link transmission from a virtual transmitter (VT) to a vehicle, which is extended to common VT as shown in Fig. \ref{fig_VT}. The state of each VT can be denoted as:
{{\begin{equation}\setlength{\abovedisplayskip}{3pt}\setlength{\belowdisplayskip}{3pt}
  \begin{array}{c}
  {\bm V}_{\left( {m,{p_m}} \right)}^{\left( k \right)} = \left\{ {{\bm r}_{\left( {m,{p_m}} \right)}^{\left( k \right)},o_{\left( {m,{p_m}} \right)}^{\left( k \right)}} \right\},\
  {\bm r}_{\left( {m,{p_m}} \right)}^{\left( k \right)} =  {\bm r}_m^{\left( k \right)} + {\bm \kappa} \left( {d_{\left( {m,{p_m}} \right)}^{\left( k \right)}-b_m,\theta _{\left( {m,{p_m}} \right)}^{\left( k \right)},\varphi _{\left( {m,{p_m}} \right)}^{\left( k \right)}} \right)
  \end{array},\label{eq_SM_5}
\end{equation}}}%
where ${\bm \kappa} \left(  \cdot  \right)$ is defined as ${\bm \kappa} \left( {d,\theta ,\varphi } \right) = d \cdot {\left( {\cos \theta \sin\varphi ,\sin \theta \sin \varphi ,\cos \varphi } \right)^T}$ that transforms the parameters of polar system to 3-D cartesian coordinate system.
  $o_{\left( {m,{p_m}} \right)}^{\left( k \right)}$   indicates the association conditions between the observation ${\bm z}_{\left( {m,{p_m}} \right)}^{\left( k \right)}$ and CVTs. In detail, if $o_{\left( {m,{p_m}} \right)}^{\left( k \right)}$ equals to 0, then ${\bm z}_{\left( {m,{p_m}} \right)}^{\left( k \right)}$ is not an observation for any legacy CVT (CVTs have been observed already). If $o_{\left( {m,{p_m}} \right)}^{\left( k \right)}$ equals to $n$, then ${\bm z}_{\left( {m,{p_m}} \right)}^{\left( k \right)}$ is exactly an observation for the $n$-th CVT.
The VTs recast from vehicle $m$ can then be denoted as $  {\bm V}_m^{\left( k \right)} = \left\{ {{\bm V}_{\left( {m,1} \right)}^{\left( k \right)},...,{\bm V}_{\left( {m,{P_m}} \right)}^{\left( k \right)}} \right\}$.

  In order to make the observation model more realistic, we also consider the false alarm (FA) and missed detection (MD) phenomenons of multi-path detection. If a multi-path measurement was not originated from any valid VT, then it is a FA (also called clutter). The number of false alarms (i.e. number of paths that do not match a valid VT) is assumed Poisson distributed with mean ${\mu _{{\text{FA}}}}$, and the distribution of false alarm measurement is described as ${{f_{{\text{FA}}}}\left( {z_{\left( {m,{p_m}} \right)}^{\left( k \right)}} \right)}$, assumed later to be a uniform distribution. The MD phenomenon means that a certain proportion of multi-path measurement (originated from valid VTs) will be missed. We denote the missed proportion as $\left( {1 - {p_d}} \right)$ so that ${{p_d}}$ is the detected probability.

\subsection{Common Virtual Transmitter Model}
\label{sec_CVT_M}
As shown in Fig. \ref{fig_VT}, a CVT is modeled based on the VTs observed simultaneously by different vehicles, which would be more precisely estimated based on multiple observations.
The CVT model is established based on the observed VTs through an affinity propagation \cite{frey2007clustering} based CVT establishment algorithm \cite{chu2020team}, and the maintenance of the CVT model is achieved in Section \ref{sec_CODA}. The state of CVTs at time $t_k$ can be denoted as:
\begin{equation}\setlength{\abovedisplayskip}{3pt}\setlength{\belowdisplayskip}{3pt}
  \begin{array}{c}
  {\bm C}_n^{\left( k \right)} = \left\{ {{\bm r}_n^{\left( {k } \right)},{\bm \varepsilon} _n^{\left( k \right)},  \rho _n^{\left( k \right)}} \right\},\
  {\bm \varepsilon} _n^{\left( k \right)} = \left\{ {{\bm \varepsilon} _{1,n}^{\left( k \right)},...,{\bm \varepsilon} _{m,n}^{\left( k \right)},...,{\bm \varepsilon} _{M,n}^{\left( k \right)}} \right\}
  \end{array},\label{eq_SM_3}
\end{equation}
\noindent where ${\bm r}_n^{\left( {k } \right)}$ is the 3-D position of the $n$-th CVT.
${\bm \varepsilon} _n^{\left( k \right)}$ is the CVT-observation association value, and ${\bm \varepsilon} _{m,n}^{\left( k \right)}$ is a single value indicating the association conditions between the CVT ${\bm C}_n^{\left( k \right)}$ and the observations from $m$-th vehicle ${\bm z}_m^{\left( k \right)}$. In detail, if ${\bm \varepsilon} _{m,n}^{\left( k \right)}$ equals to 0, then the VT ${{\bm V}_m^{\left( k \right)}}$ recast from the observation ${\bm z}_m^{\left( k \right)}$ has nothing to do with the CVT ${\bm C}_n^{\left( k \right)}$, which also means that vehicle $m$ have no observation to the CVT ${\bm C}_n^{\left( k \right)}$. If ${\bm \varepsilon} _{m,n}^{\left( k \right)}$ equals to $p_m$, then the $p_m$-th component ${\bm z}_{\left(m,p_m\right)}^{\left( k \right)}$ in ${\bm z}_m^{\left( k \right)}$ is exactly an observation to the CVT ${\bm C}_n^{\left( k \right)}$.

The value $\rho _n^{\left( k \right)}$ is the CVT-reflector association value indicating the association conditions between the CVT ${\bm C}_n^{\left (k\right)}$ and reflecting surfaces (to be introduced in Section \ref{sec_reflector_model}). In detail, if $\rho _n^{\left( k \right)} = 0$, then there is no existing reflecting surfaces associated with the CVT ${\bm C}_n^{\left (k\right)}$. If $\rho _n^{\left( k \right)} = l$, then this means that the $l$-th reflecting surface reflects the signal from the base station so that the CVT ${\bm C}_n^{\left (k\right)}$ is observed by vehicles.

 Similarly, the probabilistic distribution of ${\bm r}_n^{\left( {k } \right)}$ is also described by particles defined as  $p\left( {{\bm r}_n^{\left( k \right)}} \right) \triangleq \left\{ {p\left( {{\bm r}_n^{\left( k \right)} = {\bm r}_n^{\left( {k,j} \right)}} \right) = w_n^{\left( {k,j} \right)}} \right\}_{j = 1}^{{{\cal N}_C}}$, where ${w_n^{\left( {k,j} \right)}}$ is the weight of the particle ${{\bm r}_n^{\left( {k,j} \right)}}$, and ${{{\cal N}_C}}$ denotes the number of CVT particles.
\subsection{Model for Reflecting Surfaces}
\label{sec_reflector_model}
  This section presents the model used for reflecting surfaces, towards their estimation based on the reflective sensing and mapping shown in Section \ref{sec_environmental_awareness}. Note that we assume no prior knowledge of such surfaces or their number.
The model for each reflector (e.g. reflector ${\cal {\bm R}}_l$) is described as ${{\cal {\bm R}}_l} = \left\{ {{{\bm R}_l},{{\bm w}_l},{{\bm n}_l},{{\bm \Gamma} _l},{{\bm \gamma} _l}    } \right\}$ in this paper.
${{\bm w}_l} = {\left( {{\theta _l},{\varphi _l},{d_l}} \right)^T}$ is the basic parameter of reflector ${\cal {\bm R}}_l$, which describes the position of 3-D reflecting surface as $\sin {\varphi _l}\cos {\theta _l} \cdot x + \sin {\varphi _l}\sin {\theta _l} \cdot y + \cos {\varphi _l} \cdot z + {d_l} = 0$. ${{\bm n}_l} = {\left( {\sin {\varphi _l}\cos {\theta _l},\sin {\varphi _l}\sin {\theta _l},\cos {\varphi _l}} \right)^T}$ is the normal vector of reflector ${\cal {\bm R}}_l$, and ${{\bm \Gamma} _l} = \left\{ {\Omega _{{\theta _1}}^l,...,\Omega _{{\theta _N}}^l} \right\}$ denote the edge points of reflector ${\cal {\bm R}}_l$. ${{\bm \gamma} _l} = \left\{ {\gamma _l^{\left( 0 \right)},...,\gamma _l^{\left( k \right)},...} \right\}$ is the reflector-CVT association value indicating the association conditions between the $l$-th reflector ${\cal R}_l$ and the CVTs. In detail, if $\gamma _l^{\left( k \right)} = 0$, then the ${\cal R}_l$ is not associated with any legacy CVT. If $\gamma _l^{\left( k \right)} = n$, then this means that ${\bm C}_n^{\left (k\right)}$ is the CVT observed by certain neighboring vehicles through the multi-path signals reflected by reflector ${\cal {\bm R}}_l$.
${{\bm R}_l} = {\left( {{x_l},{y_l},{z_l}} \right)^T}$ is the position of the CVT that is symmetric with the base station about the reflector ${\cal {\bm R}}_l$, which is calculated as:
\begin{equation}\setlength{\abovedisplayskip}{3pt}\setlength{\belowdisplayskip}{3pt}
{{\bm R}_l} = \left( {I - 2{{\bm n}_l} \otimes {\bm n}_l^T} \right) \cdot {{\bm x}_{bs}} - 2{d_l}{{\bm n}_l}.\label{eq_EA_FTRL_R_l_cal}
\end{equation}

\section{Cooperative Simultaneous Localization and Mapping}
\label{sec_CO_SLAM}
The cooperative SLAM achieves cooperative multiple-vehicle localization and CVT estimation. This component firstly associates the ToA and AoA observations of each multi-path with the legacy CVTs, and then associates the reflecting surfaces with CVTs so as to utilize the sampling strategy in \cite{yin2015intelligent} to calibrate the state of CVTs for accuracy improvement. Finally, a team particle filter is utilized to estimate the position of multiple vehicles and CVTs simultaneously.

\begin{figure*}[t!]
  \centering
  \subfigure{
    \label{CODA}
    \begin{minipage}[t!]{0.52\linewidth}
      \centering
    \includegraphics[width=1\columnwidth]{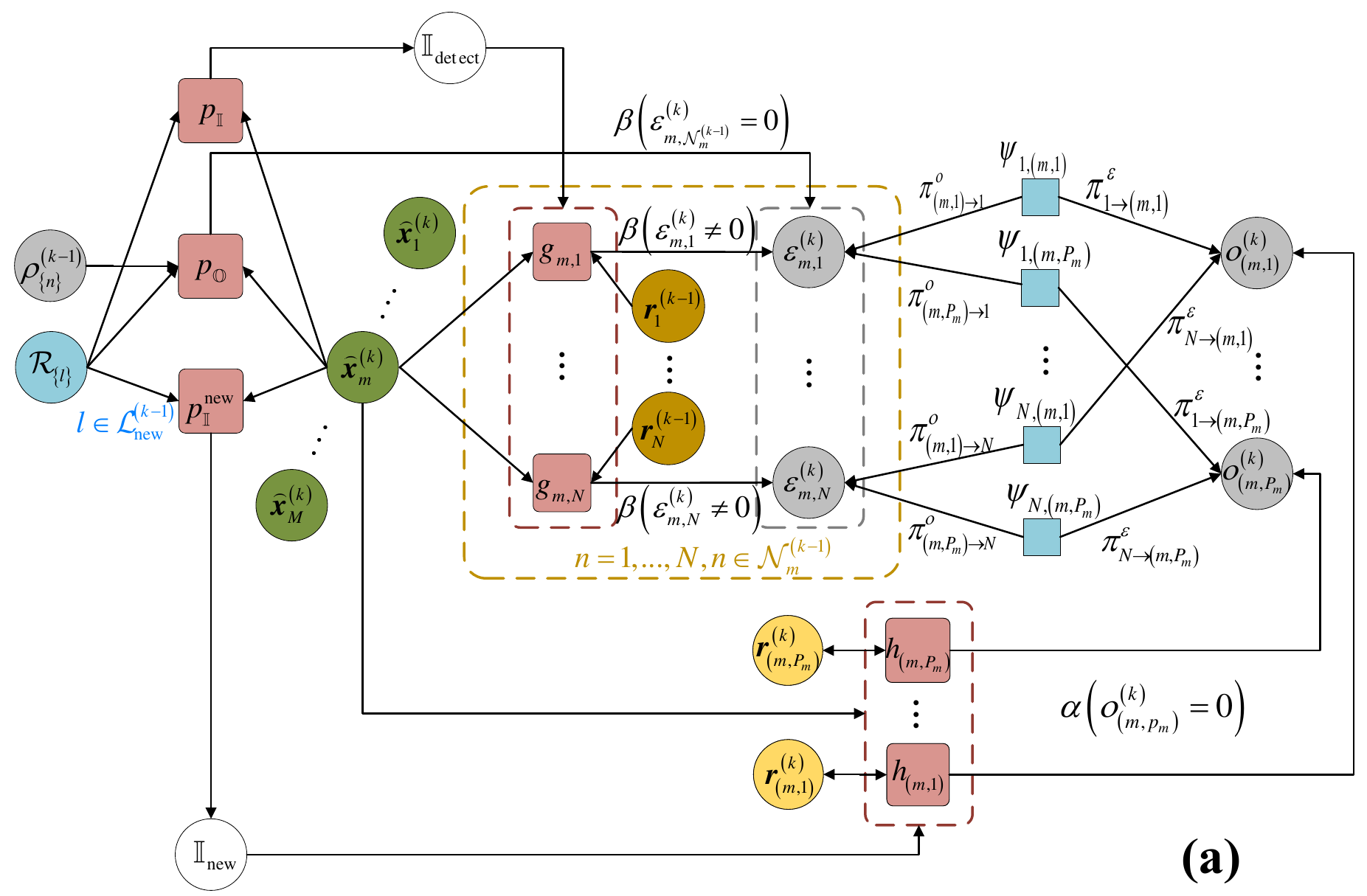}
  \end{minipage}%
  }\;
  \subfigure{
    \label{RCDA}
    \begin{minipage}[t!]{0.44\linewidth}
      \centering
      \raisebox{-1.45\height}{\includegraphics[width=1\columnwidth]{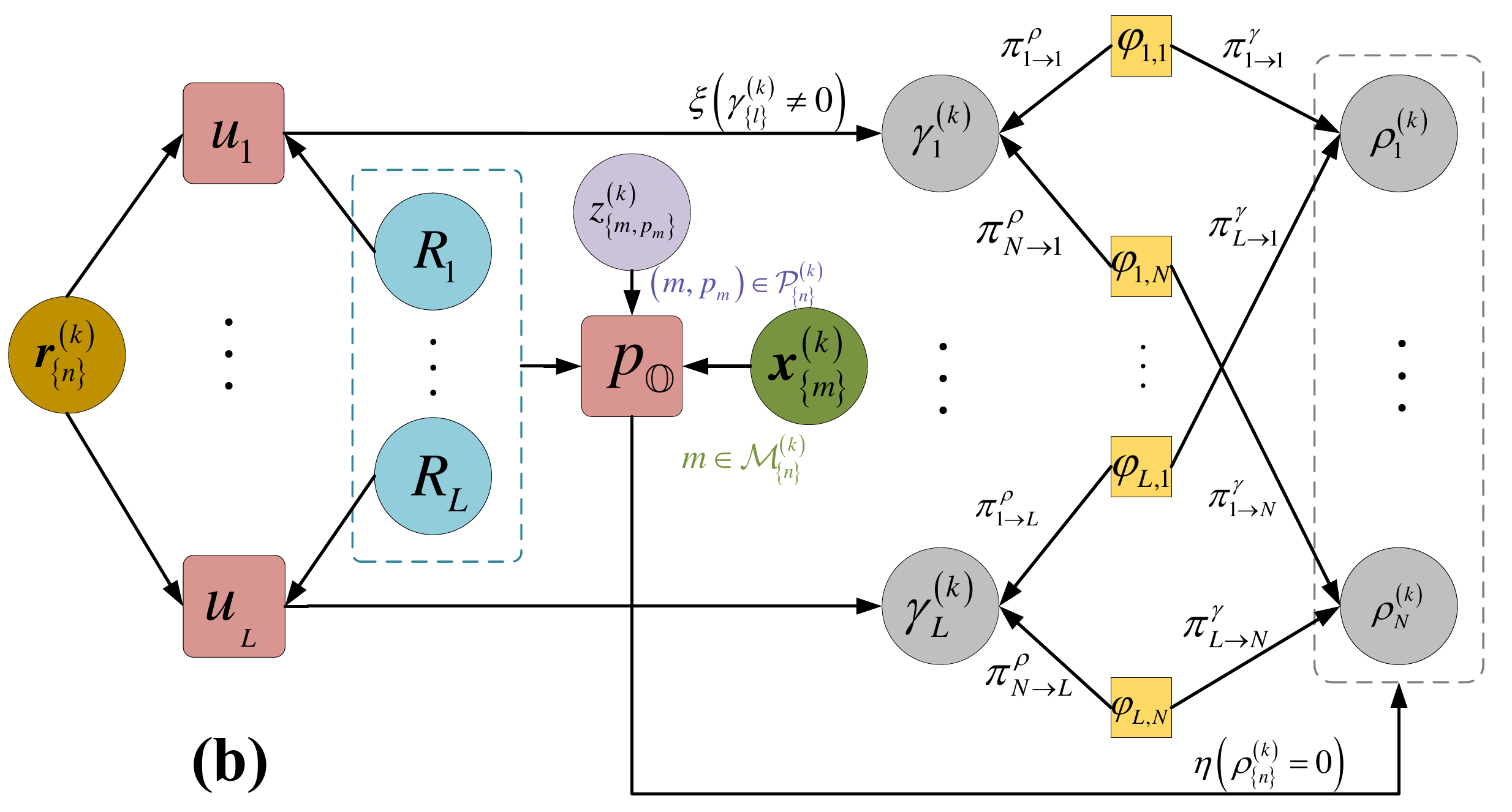}}
  \end{minipage}%
  }%
  \vspace{-1em}
  \caption{Factor graphs for data association. (a) The factor graph of CVT-observation data association for vehicle $m$, which shows the PDF of the association value between CVTs and observations (see eq. (\ref{eq_CODA_all}) in Appendix \ref{sec_CODA_appendix} for its derivation). (b) The factor graph of reflector-CVT data association, which shows the PDF of the association value between reflectors and CVTs (see eq. (\ref{eq_RCDA_all}) in Appendix \ref{sec_RCDA_appendix} for its derivation).}
  \vspace{-3ex}
\end{figure*}
\subsection{Associating ToA and AoA Observations with Common Virtual Transmitters}
\label{sec_CODA}
When the new ToA and AoA observations bring new VTs through equation (\ref{eq_SM_5}) at time $t_k$ with $N$ legacy CVTs estimated from time slot $t_0$ to $t_{k-1}$ already existing, it is necessary to associate those new VTs to the legacy CVTs to maintain the CVT model. This method allows to justify whether a new VT recast from ToA and AoA observations belongs to a certain legacy CVT or it will independently constitute a newcomer CVT. To this end, we introduce the CVT-observation data association method in this subsection based on belief propagation algorithm similar to \cite{meyer2018message}, which explores the association conditions between ToA and AoA observations as well as the legacy CVTs in a probabilistic point of view with the information of reflecting surfaces (which will be estimated through the reflective sensing and mapping component in Section \ref{sec_environmental_awareness}) considered, where the overall probability distributions are shown by the factor graph in Fig. \ref{CODA}.

The CVT-observation association value ${\bm \varepsilon} _n^{\left( k \right)}$ and the observation-CVT association value $o_{\left( {m,{p_m}} \right)}^{\left( k \right)}$ are used to associate the ToA and AoA observations with legacy CVTs. In order to include the fact that a certain multi-path observation (e.g. ${\bm z}_{\left( {m,{p_m}} \right)}^{\left( k \right)}$) can only be associated with one CVT, the global consistency constraint is introduced:
\begin{equation}\setlength{\abovedisplayskip}{3pt}\setlength{\belowdisplayskip}{3pt}
  \Phi _m^{\left( k \right)} = \prod\limits_{n \in N_m^{\left( {k - 1} \right)}} {\prod\limits_{{p_m} = 1}^{{P_m}} {{\psi _{n,\left( {m,{p_m}} \right)}}} } ,\ \ \ {\psi _{n,\left( {m,{p_m}} \right)}} = \left\{ \begin{array}{l}
    0,\ \ \ \varepsilon _{m,n}^{\left( k \right)} = {p_m},o_{\left( {m,{p_m}} \right)}^{\left( k \right)} \ne n\\
    \ \ \ \ \ \ {\rm{or}}\;o_{\left( {m,{p_m}} \right)}^{\left( k \right)} = n,\varepsilon _{m,n}^{\left( k \right)} \ne {p_m}\\
    1,\ \ \ otherwise
    \end{array} \right.\label{eq_COSLAM_constraint_def}
\end{equation}
where ${{\cal N}_m^{\left( {k - 1} \right)}}$ represents the set of CVT (indexes) that can be observed by vehicle $m$ at time slot $t_{k-1}$, which is calculated by equation (\ref{eq_CO_SLAM_CODA_N_m}) in Section \ref{sec_CO_SLAM_CODA_3}.
\subsubsection{Initial Distribution}
\label{sec_II_A_1}
As shown in Fig. \ref{CODA}, the initial belief of $o_{\left( {m,{p_m}} \right)}^{\left( k \right)}=0$ is calculated as:
{\small{\begin{equation}\setlength{\abovedisplayskip}{3pt}\setlength{\belowdisplayskip}{3pt}
    \alpha \left( {o_{\left( {m,{p_m}} \right)}^{\left( k \right)} = 0} \right)= \iint {p\left( {\overset{\lower0.5em\hbox{$\smash{\scriptscriptstyle\frown}$}}{\bm x} _m^{\left( k \right)}} \right)p\left( {{\bm r}_{\left( {m,{p_m}} \right)}^{\left( k \right)}} \right)} 
    \times {h_{\left( {m,{p_m}} \right)}}\left( {o_{\left( {m,{p_m}} \right)}^{\left( k \right)}=0\left| {\overset{\lower0.5em\hbox{$\smash{\scriptscriptstyle\frown}$}}{\bm x} _m^{\left( k \right)},{\bm r}_{\left( {m,{p_m}} \right)}^{\left( k \right)};z_{\left( {m,{p_m}} \right)}^{\left( k \right)},\mathcal{\bm R}} \right.} \right){\text{d}}\overset{\lower0.5em\hbox{$\smash{\scriptscriptstyle\frown}$}}{\bm x} _m^{\left( k \right)}{\text{d}}{\bm r}_{\left( {m,{p_m}} \right)}^{\left( k \right)}.\label{eq_Alpha_omicron_input}
\end{equation}}}
\quad Given the mean value of the number of FA and the distribution of FA measurement, the belief message ${h_{\left( {m,{p_m}} \right)}}$ in (\ref{eq_Alpha_omicron_input}) is calculated in a way similar to \cite{leitinger2019belief}.
{\small{\begin{equation}\setlength{\abovedisplayskip}{3pt}\setlength{\belowdisplayskip}{3pt}
  \begin{gathered}
    {h_{\left( {m,{p_m}} \right)}}\left( {o_{\left( {m,{p_m}} \right)}^{\left( k \right)}=0\left| {\overset{\lower0.5em\hbox{$\smash{\scriptscriptstyle\frown}$}}{\bm x} _m^{\left( k \right)},{\bm r}_{\left( {m,{p_m}} \right)}^{\left( k \right)};z_{\left( {m,{p_m}} \right)}^{\left( k \right)},\mathcal{\bm R}} \right.} \right) \hfill
     \triangleq
    \frac{{p\left( {{{\mathbbm 1}_{{\text{new}}}}\left( {z_{\left( {m,{p_m}} \right)}^{\left( k \right)}} \right);\mathcal{\bm R}} \right)p\left( {z_{\left( {m,{p_m}} \right)}^{\left( k \right)}\left| {\overset{\lower0.5em\hbox{$\smash{\scriptscriptstyle\frown}$}}{\bm x} _m^{\left( k \right)},{\bm r}_{\left( {m,{p_m}} \right)}^{\left( k \right)}} \right.} \right)}}{{{\mu _{{\text{FA}}}}{f_{{\text{FA}}}}\left( {z_{\left( {m,{p_m}} \right)}^{\left( k \right)}} \right)}}
  \end{gathered},\label{eq_h_m_pm_def}
\end{equation}}}%
where $\overset{\lower0.5em\hbox{$\smash{\scriptscriptstyle\frown}$}}{\bm x} _m^{\left( k \right)}$ is the estimated vehicle state at $t_k$  calculated as $\mathord{\buildrel{\lower3pt\hbox{$\scriptscriptstyle\frown$}}
\over {\bm r}} _m^{\left( k \right)} = {\bm r}_m^{\left( {k - 1} \right)} + \mathord{\buildrel{\lower3pt\hbox{$\scriptscriptstyle\frown$}}
\over {\bm v}} _m^{\left( k \right)}{t_\delta }$, and $\mathord{\buildrel{\lower3pt\hbox{$\scriptscriptstyle\frown$}}
\over {\bm v}} _m^{\left( k \right)}$ is the velocity at $t_k$ with Gaussian error stained.
{{The likelihood function \cite{kim20205g,8756824} is defined as:
{\small{\begin{equation}\setlength{\abovedisplayskip}{3pt}\setlength{\belowdisplayskip}{3pt}
  \begin{gathered}
    p\left( {z_{\left( {m,{p_m}} \right)}^{\left( k \right)}\left| {\overset{\lower0.5em\hbox{$\smash{\scriptscriptstyle\frown}$}}{\bm x} _m^{\left( k \right)},{\bm r}_{\left( {m,{p_m}} \right)}^{\left( k \right)}} \right.} \right) = \mathcal{G}\left( {\theta _{\left( {m,{p_m}} \right)}^{\left( k \right)} - {{\left\langle {{\bm r}_{\left( {m,{p_m}} \right)}^{\left( k \right)} - \overset{\lower0.5em\hbox{$\smash{\scriptscriptstyle\frown}$}}{\bm x} _m^{\left( k \right)}} \right\rangle }_\theta } + \alpha _m^{\left( k \right)};0,\sigma _\theta ^2 + {{\left( {\sigma _\theta ^v} \right)}^2}} \right) \hfill \\
     \cdot \mathcal{G}\left( {\varphi _{\left( {m,{p_m}} \right)}^{\left( k \right)} - {{\left\langle {{\bm r}_{\left( {m,{p_m}} \right)}^{\left( k \right)} - \overset{\lower0.5em\hbox{$\smash{\scriptscriptstyle\frown}$}}{\bm x} _m^{\left( k \right)}} \right\rangle }_\varphi };0,\sigma _\theta ^2} \right)\mathcal{G}\left( {d_{\left( {m,{p_m}} \right)}^{\left( k \right)} - \left\| {{\bm r}_{\left( {m,{p_m}} \right)}^{\left( k \right)} - \overset{\lower0.5em\hbox{$\smash{\scriptscriptstyle\frown}$}}{\bm x} _m^{\left( k \right)}} \right\| - b_m^{\left( k \right)};0,\sigma _d^2} \right) \hfill \\ 
  \end{gathered}  ,\label{eq_likelihood_def}
\end{equation}}}%
where $\mathcal{G}\left( {x;\mu ,{\sigma ^2}} \right) = \frac{1}{{\sqrt {2\pi } \sigma }}\exp \left( { - \frac{{{{\left( {x - \mu } \right)}^2}}}{{2{\sigma ^2}}}} \right)$ is the PDF of Gaussian distribution, ${\alpha _m^{\left( k \right)}}$ is the vehicle orientation at time $t_k$ with zero-mean Gaussian distributed error \cite{kim20205g} of standard deviation ${\sigma _\theta ^v}$, ${\left\langle {\bm x} \right\rangle _\theta }$ and ${\left\langle {\bm x} \right\rangle _\varphi }$ represent the azimuth angle and polar angle of vector ${\bm x}$ respectively.}} Then we utilize the information from reflecting surfaces to calculate the probability of ${{{\mathbbm 1}_{{\text{new}}}}\left( {z_{\left( {m,{p_m}} \right)}^{\left( k \right)}} \right)}$, which denotes whether ${z_{\left( {m,{p_m}} \right)}^{\left( k \right)}}$ is detected and corresponds to a new valid VT. It is calculated as:
{{\small{\begin{equation}\setlength{\abovedisplayskip}{3pt}\setlength{\belowdisplayskip}{3pt}
  p\left( {{{\mathbbm 1}_{{\text{new}}}}\left( {z_{\left( {m,{p_m}} \right)}^{\left( k \right)}} \right);\mathcal{\bm R}} \right) = \int_{l \in \mathcal{L}_{{\text{new}}}^{\left( {k - 1} \right)}} {\int {{p_{\mathbbm 1}}\left( {{{\mathbbm 1}_{\rm new}}\left| {\overset{\lower0.5em\hbox{$\smash{\scriptscriptstyle\frown}$}}{\bm x} _m^{\left( k \right)}} \right.;{\mathcal{\bm R}_l}} \right) p\left( {z_{\left( {m,{p_m}} \right)}^{\left( k \right)}\left| {\overset{\lower0.5em\hbox{$\smash{\scriptscriptstyle\frown}$}}{\bm x} _m^{\left( k \right)},{\mathcal{R}_l}} \right.} \right)p\left( {\overset{\lower0.5em\hbox{$\smash{\scriptscriptstyle\frown}$}}{\bm x} _m^{\left( k \right)}} \right) p\left( {{\mathcal{\bm R}_l}} \right){\text{d}}\overset{\lower0.5em\hbox{$\smash{\scriptscriptstyle\frown}$}}{\bm x} _m^{\left( k \right)}{\text{d}}{\mathcal{\bm R}_l}} },   \label{eq_is_a_new_detection}
\end{equation}}}}%
{{where ${p_{\mathbbm 1}}\left( {{\mathbbm 1}_{\rm new}\left| {\overset{\lower0.5em\hbox{$\smash{\scriptscriptstyle\frown}$}}{\bm x} _m^{\left( k \right)}} \right.;{\mathcal{\bm R}_l}} \right)= {p_d} \cdot {p_R}\left( {{{\mathbbm 1}_{m,l}=1}\left| {\overset{\lower0.5em\hbox{$\smash{\scriptscriptstyle\frown}$}}{\bm x} _m^{\left( k \right)};{{\cal R}_l}} \right.} \right)$ is the reflective probability with MD considered,  ${p_R}\left( {{{\mathbbm 1}_{m,l}=1}\left| {\overset{\lower0.5em\hbox{$\smash{\scriptscriptstyle\frown}$}}{\bm x} _m^{\left( k \right)};{{\cal R}_l}} \right.} \right)$ is the origin reflective probability defined by (\ref{eq_reflective_prob_defind_1}) in Section \ref{sec_RP_cal},}}
and  $\mathcal{L}_{{\text{new}}}^{\left( {k - 1} \right)} = \left\{ l \right\}\backslash {\mathcal{L}^{\left( k-1 \right)}}$ indicates the set of possible new reflectors (${\mathcal{L}^{\left( k \right)}}$ is defined in Section \ref{sec_belief_RCDA}).
Specifically, the initial belief   $\alpha \left( {o_{\left( {m,{p_m}} \right)}^{\left( k \right)}}=n \right)  \triangleq  1$.

The initial belief of ${\varepsilon _{m,n}^{\left( k \right)} = 0}$ which describes that there is no observation from vehicle $m$ associated with CVT $n$ is calculated as:
  \begin{equation}\setlength{\abovedisplayskip}{3pt}\setlength{\belowdisplayskip}{3pt}
    \beta \left( {\varepsilon _{m,n}^{\left( k \right)} = 0} \right) = \int_{l = \rho _n^{\left( {k - 1} \right)}} {\int {{p_\mathbb{O}}\left( {\mathbb{O}\left| {\overset{\lower0.5em\hbox{$\smash{\scriptscriptstyle\frown}$}}{\bm x} _m^{\left( k \right)};{\mathcal{\bm R}_l}} \right.} \right)p\left( {\overset{\lower0.5em\hbox{$\smash{\scriptscriptstyle\frown}$}}{\bm x} _m^{\left( k \right)}} \right)p\left( {{\mathcal{\bm R}_l}} \right){\text{d}}\overset{\lower0.5em\hbox{$\smash{\scriptscriptstyle\frown}$}}{\bm x} _m^{\left( k \right)}{\text{d}}{\mathcal{\bm R}_l}} }{\rm d}{\rho _n^{\left( {k - 1} \right)}},  \label{eq_no_observation_for_cvt_in_vehicle_m}
  \end{equation}
  {{where $p_{\mathbb O}\left( {{\mathbb O}\left| {\overset{\lower0.5em\hbox{$\smash{\scriptscriptstyle\frown}$}}{\bm x} _m^{\left( k \right)};{\mathcal{\bm R}_{\rho _n^{\left( {k - 1} \right)}}}} \right.} \right)= 1 - {p_d} \cdot {p_R}\left( {{{\mathbbm 1}_{m,l}=1}\left| {\overset{\lower0.5em\hbox{$\smash{\scriptscriptstyle\frown}$}}{\bm x} _m^{\left( k \right)};{{\cal R}_{\rho _n^{\left( {k - 1} \right)}}}} \right.} \right)$, and the origin  reflective probability ${p_R}\left( {{{\mathbbm 1}_{m,l}=1}\left| {\overset{\lower0.5em\hbox{$\smash{\scriptscriptstyle\frown}$}}{\bm x} _m^{\left( k \right)};{{\cal R}_{\rho _n^{\left( {k - 1} \right)}}}} \right.} \right)$ is calculated by (\ref{eq_reflective_prob_defind_1}) in Section \ref{sec_RP_cal}.}}
The initial belief of ${\varepsilon _{m,n}^{\left( k \right)} = \left( {m,{p_m}} \right)}$ is calculated as:
{\small{\begin{equation}\setlength{\abovedisplayskip}{3pt}\setlength{\belowdisplayskip}{3pt}
  \beta \left( {\varepsilon _{m,n}^{\left( k \right)}}=\left( {m,{p_m}} \right) \right) = \iint {p\left( \overset{\lower0.5em\hbox{$\smash{\scriptscriptstyle\frown}$}}{\bm x} _m^{\left( k \right)} \right)p\left( {{\bm r}_n^{\left( {k - 1} \right)}} \right)}\times {g_{m,n}}\left( {\varepsilon _{m,n}^{\left( k \right)}=\left( {m,{p_m}} \right) \left| {\overset{\lower0.5em\hbox{$\smash{\scriptscriptstyle\frown}$}}{\bm x} _m^{\left( k \right)},{\bm r}_n^{\left( {k - 1} \right)}} \right.;{\bm z}_m^{\left( k \right)}},{\cal {\bm R}} \right){\text{d}}\overset{\lower0.5em\hbox{$\smash{\scriptscriptstyle\frown}$}}{\bm x} _m^{\left( k \right)}{\text{d}}{\bm r}_n^{\left( {k - 1} \right)}. \label{eq_Beta_sigma_input}
\end{equation}}}
The belief message ${g_{m,n}}$ is calculated as:
{\small{\begin{equation}
    \begin{gathered}
      {g_{m,n}}\left( {\varepsilon _{m,n}^{\left( k \right)}=\left( {m,{p_m}} \right) \left| {\overset{\lower0.5em\hbox{$\smash{\scriptscriptstyle\frown}$}}{\bm x} _m^{\left( k \right)},{\bm r}_n^{\left( {k - 1} \right)}} \right.;z_m^{\left( k \right)},\mathcal{\bm R}} \right) \hfill
       \triangleq
      \frac{{p\left( {{{\mathbbm 1}_{{\text{observe}}}}\left( {z_{\left( {m,{p_m}} \right)}^{\left( k \right)}} \right);\mathcal{\bm R}} \right)p\left( {z_{\left( {m,{p_m}} \right)}^{\left( k \right)}\left| {\overset{\lower0.5em\hbox{$\smash{\scriptscriptstyle\frown}$}}{\bm x} _m^{\left( k \right)},} \right.{\bm r}_n^{\left( {k - 1} \right)}} \right)}}{{{\mu _{{\text{FA}}}}{f_{{\text{FA}}}}\left( {z_{\left( {m,{p_m}} \right)}^{\left( k \right)}} \right)}}
    \end{gathered}, \label{eq_g_m_n_def}
  \end{equation}}}%
where ${{{\mathbbm 1}_{{\text{observe}}}}\left( {z_{\left( {m,{p_m}} \right)}^{\left( k \right)}} \right)}$ denotes whether ${z_{\left( {m,{p_m}} \right)}^{\left( k \right)}}$ is detected and corresponds to a new or legacy valid VT. We define the probability of it as observing probability which is calculated as:
{{\small{  \begin{equation}
    p\left( {{{\mathbbm 1}_{{\text{observe}}}}\left( {z_{\left( {m,{p_m}} \right)}^{\left( k \right)}} \right);\mathcal{\cal R}} \right) = \int_{l \in \left\{ l \right\}} {\int {{p_{\mathbbm 1}}\left( {{{\mathbbm 1}_{{\text{observe}}}}\left| {\overset{\lower0.5em\hbox{$\smash{\scriptscriptstyle\frown}$}}{\bm x} _m^{\left( k \right)}} \right.;{\mathcal{\bm R}_l}} \right)p\left( {z_{\left( {m,{p_m}} \right)}^{\left( k \right)}\left| {\overset{\lower0.5em\hbox{$\smash{\scriptscriptstyle\frown}$}}{\bm x} _m^{\left( k \right)},{\mathcal{R}_l}} \right.} \right)p\left( {\overset{\lower0.5em\hbox{$\smash{\scriptscriptstyle\frown}$}}{\bm x} _m^{\left( k \right)}} \right)p\left( {{\mathcal{\bm R}_l}} \right){\text{d}}\overset{\lower0.5em\hbox{$\smash{\scriptscriptstyle\frown}$}}{\bm x} _m^{\left( k \right)}{\text{d}}{\mathcal{\bm R}_l}} } ,\label{eq_cal_1_observe}
  \end{equation}}}}%
{{where ${p_{\mathbbm 1}}\left( {{{\mathbbm 1}_{{\text{observe}}}}\left| {\overset{\lower0.5em\hbox{$\smash{\scriptscriptstyle\frown}$}}{\bm x} _m^{\left( k \right)}} \right.;{\mathcal{\bm R}_l}} \right)= {p_d} \cdot {p_R}\left( {{{\mathbbm 1}_{m,l}=1}\left| {\overset{\lower0.5em\hbox{$\smash{\scriptscriptstyle\frown}$}}{\bm x} _m^{\left( k \right)};{{\cal R}_l}} \right.} \right)$ is the reflective probability with MD considered, and ${p_R}\left( {{{\mathbbm 1}_{m,l}=1}\left| {\overset{\lower0.5em\hbox{$\smash{\scriptscriptstyle\frown}$}}{\bm x} _m^{\left( k \right)};{{\cal R}_l}} \right.} \right)$ is  defined by (\ref{eq_reflective_prob_defind_1}) in Section \ref{sec_RP_cal}.}}

\subsubsection{Belief Message Propagation and Association}
\label{sec_CO_SLAM_CODA_3}
The belief message $\pi _{n \to \left( {m,{p_m}} \right)}^{\varepsilon \left[ {iter} \right]}$ from CVT-observation association value to observation-CVT association value
can be obtain iteratively by belief propagation algorithm in \cite{meyer2018message}.
The probability that the observation of multi-path ${\left( {m,{p_m}} \right)}$ is associated with the CVT ${\bm C}_n^{\left(k\right)}$ is calculated as:
\begin{equation}\setlength{\abovedisplayskip}{3pt}\setlength{\belowdisplayskip}{3pt}
  \begin{gathered}
    \overset{\lower0.5em\hbox{$\smash{\scriptscriptstyle\frown}$}}{p} \left( {o_{\left( {m,{p_m}} \right)}^{\left( k \right)} = n} \right) \hfill
     = \frac{{\alpha \left( {o_{\left( {m,{p_m}} \right)}^{\left( k \right)} = n} \right)\pi _{n \to \left( {m,{p_m}} \right)}^{\varepsilon \left[ {ITER} \right]}}}{{\alpha \left( {o_{\left( {m,{p_m}} \right)}^{\left( k \right)} = 0} \right) + \sum\limits_{n'} {\alpha \left( {o_{\left( {m,{p_m}} \right)}^{\left( k \right)} = n'} \right)\pi _{n' \to \left( {m,{p_m}} \right)}^{\varepsilon \left[ {ITER} \right]}} }}.
  \end{gathered}  \label{eq_CODA_output}
\end{equation}
{{The CVT (index) $\mathord{\buildrel{\lower3pt\hbox{$\scriptscriptstyle\frown$}}
\over o} _{\left( {m,{p_m}} \right)}^{\left( k \right)}$ that is associated with the observation of multi-path ${\left( {m,{p_m}} \right)}$ is then calculated following the most probable principle \cite{charniak1992dynamic} in order to improve the efficiency of TCSE:}}
\begin{equation}\setlength{\abovedisplayskip}{3pt}\setlength{\belowdisplayskip}{3pt}
  \mathord{\buildrel{\lower3pt\hbox{$\scriptscriptstyle\frown$}}
\over o} _{\left( {m,{p_m}} \right)}^{\left( k \right)} = \mathop {\arg \max }\limits_n \overset{\lower0.5em\hbox{$\smash{\scriptscriptstyle\frown}$}}{p} \left( {o_{\left( {m,{p_m}} \right)}^{\left( k \right)} = n} \right).\label{eq_CODA_choose}
\end{equation}

Thus the association between the observations and the legacy CVTs can be executed as:
\begin{itemize}
  \item  {\em Mitigation of false alarms:} If the probability $ p\left( {{{\mathbbm 1}_{{\text{observe}}}}\left( {z_{\left( {m,{p_m}} \right)}^{\left( k \right)}} \right);\mathcal{\cal R}} \right)$ is smaller than a FA threshold $\delta_{\rm FA}$, then the observation $z_{\left( {m,{p_m}} \right)}^{\left( k \right)}$ will be seen as a FA and discarded.
  \item {\em Observation association:} An observation $z_{\left( {m,{p_m}} \right)}^{\left( k \right)}$ will be  associated to a legacy CVT according to (\ref{eq_CODA_choose}) if $\mathord{\buildrel{\lower3pt\hbox{$\scriptscriptstyle\frown$}}
  \over o} _{\left( {m,{p_m}} \right)}^{\left( k \right)}  \ne 0$, otherwise its corresponding VT calculated by (\ref{eq_SM_5}) will be regarded as a standalone CVT.
  \item {\em Discarding of empty CVT:} A legacy CVT is discarded if it has no associated observation.
\end{itemize}

  We utilize ${\cal N}_m^{\left( {k} \right)}$ to denote the set of CVT (indexes) that are associated with observations from vehicle $m$. We utilize ${\cal P}_n^{\left( k \right)}$ to denote the set of multi-path observation (indexes) that are associated with CVT ${\bm C}_n^{\left(k\right)}$, and utilize ${\cal M}_n^{\left( k \right)}$ to denote the set of vehicle (indexes) that have multi-path observations associated with CVT ${\bm C}_n^{\left(k\right)}$, which are calculated respectively as:
  \begin{equation}\setlength{\abovedisplayskip}{3pt}\setlength{\belowdisplayskip}{3pt}
    {\cal N}_m^{\left( k \right)} = \mathop  \cup \limits_{{p_m}} \left\{ {\mathord{\buildrel{\lower3pt\hbox{$\scriptscriptstyle\frown$}}
    \over o} _{\left( {m,{p_m}} \right)}^{\left( k \right)}} \right\},{\cal P}_n^{\left( k \right)} = \left\{ {\left( {m,{p_m}} \right)\left| {\mathord{\buildrel{\lower3pt\hbox{$\scriptscriptstyle\frown$}}
    \over o} _{\left( {m,{p_m}} \right)}^{\left( k \right)} = n} \right.} \right\},{\cal M}_n^{\left( k \right)} = \left\{ {m\left| {\overset{\lower0.5em\hbox{$\smash{\scriptscriptstyle\frown}$}}{o} _{\left( {m,{p_m}} \right)}^{\left( k \right)} = n} \right.} \right\}.\label{eq_CO_SLAM_CODA_N_m}
  \end{equation}
\subsection{Associating Reflecting Surfaces with Common Virtual Transmitters}
\label{sec_CO_SLAM_RCDA}
The information of reflecting surfaces estimated from the reflective sensing and mapping in Section \ref{sec_environmental_awareness} can be utilized by the cooperative SLAM to further improve the accuracy of CVT estimation.
To this end, it is necessary to explore the association conditions between the reflecting surfaces and CVTs to allow the sampling strategy introduced in Section \ref{sec_CO_SLAM_IS} to deliver the information of reflecting surfaces to CVTs. So we introduce the reflector-CVT data association method in this subsection with the factor graph shown in Fig. \ref{RCDA}.
The reflector-CVT association value $\gamma _l^{\left( k \right)}$ and CVT-reflector association value $\rho _n^{\left( k \right)}$ are utilized to associate the reflecting surfaces with legacy CVTs. Similarly, the global consistency constraint is defined as:
\begin{equation}\setlength{\abovedisplayskip}{3pt}\setlength{\belowdisplayskip}{3pt}
  \begin{array}{c}
{\Psi ^{\left( k \right)}}  = \prod\limits_{l = 1}^L {\prod\limits_{n = 1}^N}  {{{\varphi _{l,n}}}  ,\ \ \ 
{\varphi _{l,n}}}  = \left\{ \begin{array}{l}
0,{\rm{\ \ \ }}\gamma _l^{\left( k \right)} = n,\rho _n^{\left( k \right)} \ne l \ {\rm or}\ \rho _n^{\left( k \right)} = l,\gamma _l^{\left( k \right)} \ne n\\
1,{\rm{\ \ \ \ }}otherwise
\end{array} \right.
\end{array}.\label{eq_CO_SLAM_RCDA_constrain}
\end{equation}
\subsubsection{Initial Distribution}
\label{sec_coslam_rcda_id}
The initial belief of ${\gamma _l^{\left( k \right)}} = n$ is calculated as:
  \begin{equation}\setlength{\abovedisplayskip}{3pt}\setlength{\belowdisplayskip}{3pt}
    \begin{gathered}
    \xi \left( {\gamma _l^{\left( k \right)}}=n \right) \hfill
    {\text{ = }}\iint {u_l\left( {\gamma _l^{\left( k \right)}=n\left| {{{\bm r}_{\left\{ n \right\}}^{\left( k \right)}},{{ {\cal {\bm R}}}_l}} \right.} \right)}p\left( {{{\cal {\bm R}}_l}} \right) p\left( {{{\bm r}_{\left\{ n \right\}}^{\left( k \right)}}} \right){\text{d}}{{\bm r}_{\left\{ n \right\}}^{\left( k \right)}}{\text{d}}{{ \cal {\bm R}}_l},
  \end{gathered} \label{eq_Xi_gamma_input}
  \end{equation}
where the belief message $u_l$ is defined as:
\begin{equation}\setlength{\abovedisplayskip}{3pt}\setlength{\belowdisplayskip}{3pt}
  u_l\left( {\gamma _l^{\left( k \right)}=n\left| {{{\bm r}_{\left\{ n \right\}}^{\left( k \right)}},{{\cal  {\bm R}}_l}} \right.} \right) \triangleq
  p\left( {\gamma _l^{\left( k \right)}\left| {{\bm C}_n^{\left( k \right)},{{\cal {\bm R}}_l}} \right.} \right). \label{eq_u_l_def}
\end{equation}
Specially, we define the belief $\xi \left( {\gamma _l^{\left( k \right)} = 0} \right) \triangleq 1$.
The initial belief of ${\rho _n^{\left( k \right)} = 0}$ is calculated as:
{{\small{\begin{equation}
    \eta \left( {\rho _n^{\left( k \right)} = 0} \right) = \iiint_{\begin{subarray}{l}
      \left( {m,{p_m}} \right) \in \mathcal{P}_n^{\left( k \right)} ,\\
      m \in \mathcal{M}_n^{\left( k \right)},l \in \left\{ l \right\}
    \end{subarray}}  {p_{\mathbb O}\left( {{\mathbb O}\left| {{\bm x}_m^{\left( k \right)};{\mathcal{\bm R}_l}} \right.} \right)p\left( {z_{\left( {m,{p_m}} \right)}^{\left( k \right)}\left| {{\bm x}_m^{\left( k \right)},{\mathcal{R}_l}} \right.} \right)p\left( {{\bm x}_m^{\left( k \right)}} \right)p\left( {{{\cal{\bm R}}_l}} \right){\text{d}}z_{\left( {m,{p_m}} \right)}^{\left( k \right)}{\text{d}}{\bm x}_m^{\left( k \right)}{\text{d}}{{\cal {\bm R}}_l}},\label{eq_Y_rho_0}
  \end{equation}}}}%
{{where $p_{\mathbb O}\left( {{\mathbb O}\left| {{\bm x}_m^{\left( k \right)};{\mathcal{\bm R}_l}} \right.} \right)= 1 - {p_d} \cdot {p_R}\left( {{{\mathbbm 1}_{m,l}=1}\left| {{\bm x} _m^{\left( k \right)};{{\cal R}_l}} \right.} \right)$, and ${p_R}\left( {{{\mathbbm 1}_{m,l}=1}\left| {{\bm x} _m^{\left( k \right)};{{\cal R}_l}} \right.} \right)$ is the origin  reflective probability calculated by (\ref{eq_reflective_prob_defind_1}) in Section \ref{sec_RP_cal}.}}
Specifically, the initial belief $\eta \left( {\rho _n^{\left( k \right)} = l} \right) \triangleq 1$.

\subsubsection{Belief Message Propagation and Association}
\label{sec_belief_RCDA}
The belief message $\pi _{l \to n}^{\gamma \left[ {iter} \right]}$ from the reflector-CVT association value to CVT-reflector association value
 can be obtained iteratively by the belief propagation algorithm in \cite{meyer2018message}.
The probability that the CVT ${\bm C}_n^{\left(k\right)}$ is associated to reflector ${\cal {\bm R}}_l$ is then calculated as:
\begin{equation}\setlength{\abovedisplayskip}{3pt}\setlength{\belowdisplayskip}{3pt}
  \begin{gathered}
    {\mathord{\buildrel{\lower3pt\hbox{$\scriptscriptstyle\frown$}}
\over p} }\left( {\rho _n^{\left( k \right)} = l} \right) \hfill
     = \frac{{\eta \left( {\rho _n^{\left( k \right)} = l} \right)\pi _{l \to n}^{\gamma \left[ {ITER} \right]}}}{{\eta \left( {\rho _n^{\left( k \right)} = 0} \right) + \sum\limits_{l'} {\eta \left( {\rho _n^{\left( k \right)} = l'} \right)\pi _{l' \to n}^{\gamma \left[ {ITER} \right]}} }} \hfill \\
  \end{gathered}.  \label{eq_RCDA_output}
\end{equation}
{{So the reflector (index) that associates with the CVT ${\bm C}_n^{\left(k\right)}$ is calculated following the most probable principle \cite{charniak1992dynamic}  in order to improve the efficiency of TCSE:}}
\begin{equation}\setlength{\abovedisplayskip}{3pt}\setlength{\belowdisplayskip}{3pt}
  \mathord{\buildrel{\lower3pt\hbox{$\scriptscriptstyle\frown$}}
\over \rho } _n^{\left( k \right)} = \mathop {\arg \max }\limits_l \overset{\lower0.5em\hbox{$\smash{\scriptscriptstyle\frown}$}}{p} \left( {\rho _n^{\left( k \right)} = l} \right).\label{eq_RCDA_choose}
\end{equation}
  Thus we define the set of reflectors associated with legacy CVTs as ${\mathcal{L}^{\left( k \right)}} = \bigcup\limits_n {\overset{\lower0.5em\hbox{$\smash{\scriptscriptstyle\frown}$}}{\rho } _n^{\left( k \right)}} $.

\subsection{Sampling Strategy for Common Virtual Transmitter Particles}
\label{sec_CO_SLAM_IS}
In this section, a recently introduced sampling strategy \cite{yin2015intelligent} is used to deliver the information of reflecting surfaces (estimated from reflective sensing and mapping in Section \ref{sec_environmental_awareness}) to CVTs. The purpose of this method is to improve the particle sampling accuracy for CVTs. In the particle based estimation method like particle filter, the particle sampling process refers to approximating a certain probability distribution by sampling a finite number of particles   based on pre-modeled uncertainties (e.g. Gaussian uncertainty) \cite{van2001unscented,Siciliano2016Robotics,gustafsson2002particle}. However, if some information about the probability distribution (like the information of reflecting surfaces in this paper) is already known, the particles can be drawn more efficiently. So this section allows to deliver the information of reflecting surfaces to the CVT estimation in the CVT particle sampling process.
In detail, the information from the reflectors can help to select the CVT particles more reasonably so as to calibrate the position of the CVTs for vehicular positioning accuracy improvement.

\subsubsection{Weight updating}
\label{sec_CO_SLAM_IS_1}
For a CVT (e.g. the CVT ${\bm C}_n^{\left(k\right)}$), the information of reflecting surfaces and CVT-reflector association value $\mathord{\buildrel{\lower3pt\hbox{$\scriptscriptstyle\frown$}}
\over p} \left( {\rho _n^{\left( k \right)} = l} \right)$ are used to update the weight of its particles:
\begin{equation}\setlength{\abovedisplayskip}{3pt}\setlength{\belowdisplayskip}{3pt}
  \begin{array}{l}
  p\left( {{\bm r}_n^{\left( k \right)}} \right) =
  \iint {p\left( {{\bm r}_n^{\left( k \right)}\left| {\rho _n^{\left( k \right)},{{\bm R}_{\left\{ l \right\}}}} \right.} \right)\mathord{\buildrel{\lower3pt\hbox{$\scriptscriptstyle\frown$}}
  \over p} \left( {\rho _n^{\left( k \right)} = l} \right)p\left( {{\bm R}_{\left\{ l \right\}}} \right){\text{d}}\rho _n^{\left( k \right)}{\text{d}}{{\bm R}_{\left\{ l \right\}}}}
  \end{array}.\label{eq_IS_weight_update}
\end{equation}
\subsubsection{Particle crossover and mutation}
The particles updated through equation (\ref{eq_IS_weight_update}) are then divided into higher-weighted particles and lower-weighted particles. Then crossover and mutation operations are adopted to sample the particles. In detail, the crossover operation makes all the lower-weighted particles get closer to the higher-weighted particles. The mutation operation generates new higher-weighted particles with a mutation probability $p_M$. The crossover and mutation operations can be summarized as:

{\em {\textbf{Step 1:}}} Calculate the threshold to distinguish higher-weighted particles and lower-weighted particles as $  N_{{\rm eff}}^{k,n} = \left\lfloor {{1 \mathord{\left/
   {\vphantom {1 {\sum\limits_{j = 1}^{{N_C}} {{{\left( {w_{{C_u}}^{\left( j \right)}} \right)}^2}} }}} \right.
   \kern-\nulldelimiterspace} {\sum\limits_{j = 1}^{{{\cal N}_C}} {{{\left( {w_{{n}}^{\left( k,j \right)}} \right)}^2}} }}} \right\rfloor$.

{\em {\textbf{Step 2}}}: Sort the particles in the descending order of the weights calculated by (\ref{eq_IS_weight_update}). The particles ranked before the $N_{{\rm eff}}^{k,n}$-th particle are defined as higher-weighted particles denoted as $\left\{ {{\bm r}_n^{\left( {k,{j_H}} \right)},w_n^{\left( {k,{j_H}} \right)}} \right\}$, and the particles ranked after the $N_{{\rm eff}}^{k,n}$-th particle are defined as lower-weighted particles denoted as $\left\{ {{\bm r}_n^{\left( {k,{j_L}} \right)},w_n^{\left( {k,{j_L}} \right)}} \right\}$.

{\em {\textbf{Step 3}}}: Update the particles in a joint crossover and mutation manner:
\begin{equation}\setlength{\abovedisplayskip}{3pt}\setlength{\belowdisplayskip}{3pt}
  {\bm r}_n^{\left( {k,{j_L}} \right)} = \left\{ \begin{array}{l}
  \alpha_C \left( {2{\bm r}_n^{\left( {k,{j_H}} \right)} - {\bm r}_n^{\left( {k,{j_L}} \right)}} \right) + \left( {1 - \alpha_C } \right){\bm r}_n^{\left( {k,{j_H}} \right)},
  rand\left( {0,1} \right) \le {p_M}\\
  \alpha_C {\bm r}_n^{\left( {k,{j_L}} \right)} + \left( {1 - \alpha_C } \right){\bm r}_n^{\left( {k,{j_H}} \right)},rand\left( {0,1} \right) > {p_M}
  \end{array} \right.,\label{eq_IS_cm_2}
\end{equation}
where $\alpha_C \in \left( {0,1} \right)$ determines how much information from ${\bm r}_n^{\left( {k,{j_H}} \right)}$ is supposed to be transferred to ${\bm r}_n^{\left( {k,{j_L}} \right)}$ in the crossover operation, and $p_M$ is the mutation probability indicating the possibility that a lower-weighted particle is mutated into a new higher-weighted particle.

\subsection{Team Particle Filter}

Team particle filter can be seen as  an extension of classical particle filter \cite{van2001unscented,Siciliano2016Robotics,gustafsson2002particle} and is here proposed as a way to estimate the positions of multiple vehicles and CVTs simultaneously (such position estimates are modeled by particles) using stochastic batch iteration \cite{chu2021vehicle}, {{which utilizes the information of reflecting surfaces together with the sampling strategy in Section \ref{sec_CO_SLAM_IS}.}}

In each time slot (e.g. time slot $t_k$), the team particle filter firstly samples the particles of each CVT based on the information of reflecting surfaces and the particles of each vehicle based on its velocity information. Then it divides the vehicle particles as well as the CVT particles into stochastic batches, and updates the position of CVTs and multiple vehicles iteratively. The team particle filter is described as follows:

\subsubsection{Particle sampling}
Since the position of CVTs are static over time, the particles of each CVT (e.g. CVT ${\bm C}_n^{\left(k\right)}$) are inherited from its previous probability distribution as:
\begin{equation}\setlength{\abovedisplayskip}{3pt}\setlength{\belowdisplayskip}{3pt}
  {\bm r}_n^{\left( {k,j} \right)} \sim p\left( {{\bm r}_n^{\left( {k,j} \right)}\left| {{\bm r}_n^{\left( {k - 1,j} \right)}} \right.} \right) = \delta \left( {{\bm r}_n^{\left( {k,j} \right)} - {\bm r}_n^{\left( {k - 1,j} \right)}} \right),\label{eq_TP_1}
\end{equation}
where $\delta \left(  \cdot  \right)$ denotes the impulse function. Then the particles of each CVT are updated by the sampling strategy described in Section \ref{sec_CO_SLAM_IS}.

The particles of vehicle $m$ are drawn by its velocity ${\bm v}_m^{\left( k \right)}$ described in Section \ref{sec_II_A_1}:
\begin{equation}\setlength{\abovedisplayskip}{3pt}\setlength{\belowdisplayskip}{3pt}
  {\bm r}_m^{\left( {k,i} \right)} \sim p\left( {{\bm r}_m^{\left( {k,i} \right)}\left| {{\bm r}_m^{\left( {k - 1,i} \right)},{\bm v}_m^{\left( k \right)}} \right.} \right).\label{eq_TP_2}
\end{equation}

\subsubsection{Joint updating}
In each iteration, stochastic batches are chosen from the particles of each vehicle and each CVT, so the updating process in each iteration (e.g. iteration $b$) are summarized as:

{\em \textbf{Step 1:}} Update the particles chosen in the iteration $b$ for each CVT (e.g. CVT ${\bm C}_n^{\left(k\right)}$) as:
{\small{\begin{equation}\setlength{\abovedisplayskip}{3pt}\setlength{\belowdisplayskip}{3pt}
  w_n^{\left( {k,j_{n,b}} \right)} = w_n^{\left( {k - 1,j_{n,b}} \right)}  p\left( {{\bm z}_{{\cal P}_n^{\left( k \right)}}^{\left( k \right)}\left| {{\bm r}_n^{\left( k \right)}} \right.} \right) =w_n^{\left( {k - 1,j_{n,b}} \right)}
  \prod\limits_{\left( {m,{p_m}} \right) \in {\cal P}_n^{\left( k \right)}} {\int {p\left( {{\bm z}_{\left( {m,{p_m}} \right)}^{\left( k \right)}\left| {{\bm r}_n^{\left( k \right)},{\bm r}_m^{\left( k \right)}} \right.} \right)p\left( {{\bm r}_m^{\left( k \right)}} \right)} {\rm{d}}{\bm r}_m^{\left( k \right)}},\label{eq_TP_3}
\end{equation}}}%
where $j_{n,b}$ denotes the index of particles belonging to the set of particles chosen in the $b$-th iteration for CVT ${\bm C}_n^{\left(k\right)}$, and ${\cal P}_n^{\left( k \right)}$ is the set of multi-path observation (indexes) that is associated with CVT ${{\bm C}}_n^{\left(k\right)}$ calculated in (\ref{eq_CO_SLAM_CODA_N_m}).

{\em \textbf{Step 2:}} Update the particles chosen in the iteration $b$ for each vehicle (e.g. vehicle $m$) as:
{\small{\begin{equation}\setlength{\abovedisplayskip}{3pt}\setlength{\belowdisplayskip}{3pt}
  w_m^{\left( {k,i_{m,b}} \right)} = w_m^{\left( {k - 1,i_{m,b}} \right)} p\left( {{\bm z}_m^{\left( k \right)}\left| {{\bm r}_m^{\left( k \right)}} \right.} \right) = w_m^{\left( {k - 1,i_{m,b}} \right)}
  \prod\limits_{\begin{subarray}{c}
{p_m} = 1,
n = \mathord{\buildrel{\lower3pt\hbox{$\scriptscriptstyle\frown$}}
\over o} _{\left( {m,{p_m}} \right)}^{\left( k \right)}
\end{subarray}}^{P_m} {\int {p\left( {{\bm z}_{\left( {m,{p_m}} \right)}^{\left( k \right)}\left| {{\bm r}_m^{\left( k \right)},{\bm r}_n^{\left( k \right)}} \right.} \right)p\left( {{\bm r}_n^{\left( k \right)}} \right)} {\rm{d}}{\bm r}_n^{\left( k \right)}},
\label{eq_TP_4}  
\end{equation}}}%
where $i_{m,b}$ denotes the index of particles belonging to the set of particles chosen in the $b$-th iteration for vehicle $m$, and ${\mathord{\buildrel{\lower3pt\hbox{$\scriptscriptstyle\frown$}}
\over o} _{\left( {m,{p_m}} \right)}^{\left( k \right)}}$ is the CVT (index) that is associated with the observation of multi-path ${\left( {m,{p_m}} \right)}$ calculated in (\ref{eq_CODA_choose}).

\section{Reflective Sensing and Mapping}
\label{sec_environmental_awareness}

The reflective sensing and mapping component estimates the position and edge of the reflecting surfaces by extracting information from the position estimations of multiple vehicles and CVTs in cooperative SLAM. This component also in return provides the information of reflecting surfaces for cooperative SLAM and radio geometrization for positioning accuracy improvement. In detail, this component firstly extracts the reflecting elements in Section \ref{sec_EA_1} as shown in Fig. \ref{fig_reflecting_element_edge_estimation}. Then an online learning approach based on FTRL \cite{duchi2011adaptive} is introduced to estimate the position and edge of the reflecting surfaces in Section \ref{sec_EA_FTRL_EE}, where the reflective probability\footnotemark[2] ${p_R}\left( {{\mathbbm 1}_{m,l}\left| {\bm x}_m \right.;{\mathcal{\bm R}_l}} \right)$ is calculated based on the ray-crossing algorithm \cite{sutherland1974characterization}.

\footnotetext[2] {The reflective probability ${p_R}\left( {{\mathbbm 1}_{m,l}\left| {\bm x}_m \right.;{\mathcal{\bm R}_l}} \right)$ indicates the probability that whether a vehicle locates at position ${\bm r}_m$ can receive the signals reflected by reflector ${\cal {\bm R}}_l$.}

\subsection{Reflecting Element Extraction}
\label{sec_EA_1}
\begin{figure}[!t]
  \setlength{\abovecaptionskip}{-0.2cm}
  \centering
  \includegraphics[width=0.55\columnwidth]{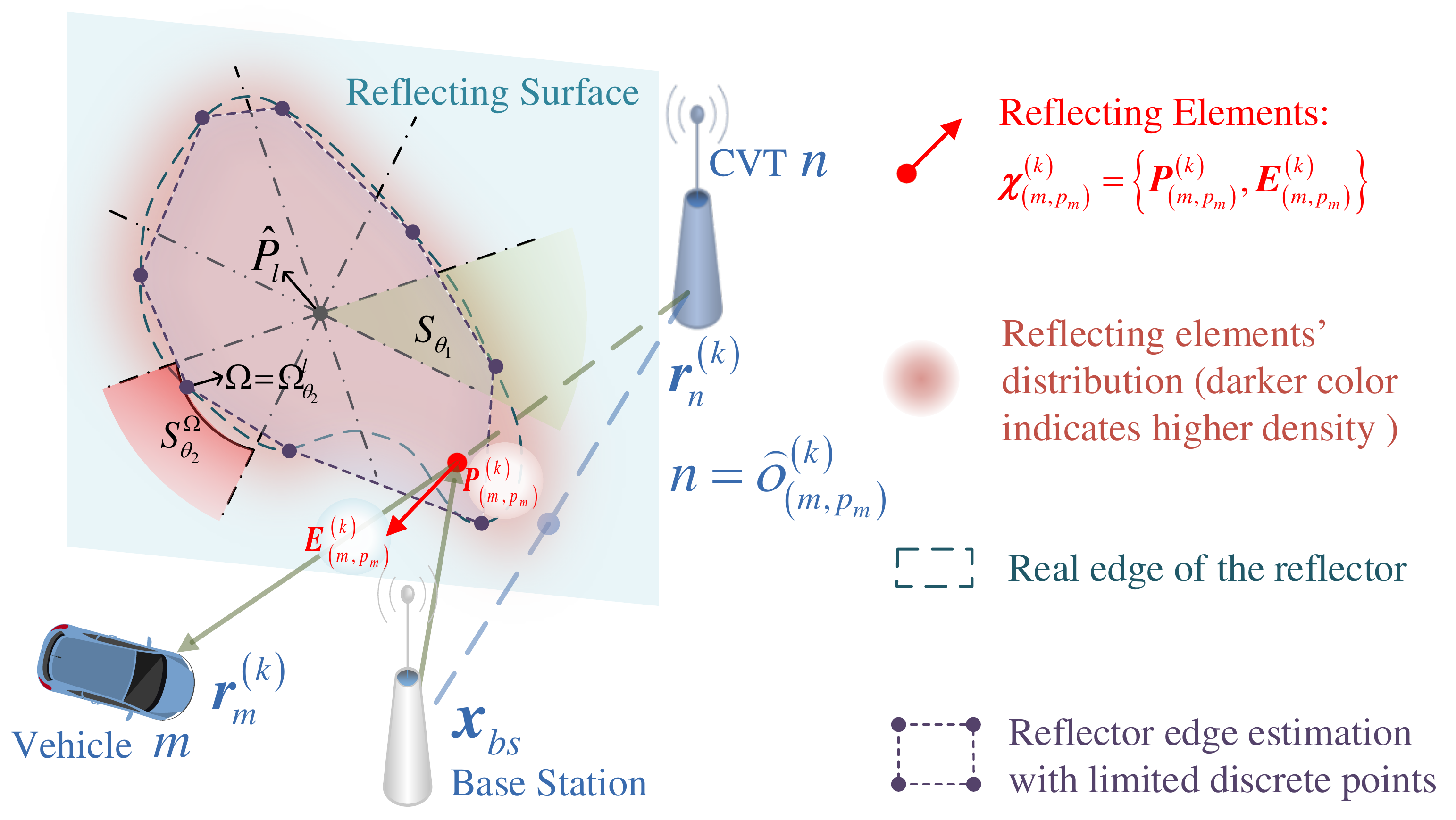}
  \caption{Reflecting elements and edge estimation. This figure shows the reflecting elements for the $p_m$-th path observed by vehicle $m$, and also shows the edge estimation of a reflector using 8 discrete points (${\cal V}_l^\theta = 8$).
  }
  \label{fig_reflecting_element_edge_estimation}
  \vspace{-4ex}
\end{figure}

The reflecting elements are extracted from each multi-path observation as shown in Fig. \ref{fig_reflecting_element_edge_estimation}. For the multi-path indexed by ${\left( {m,{p_m}} \right)}$, the corresponding reflecting element includes: 1) the position of its reflecting point, and 2) the normal vector of its reflecting surface, which can be denoted as ${\bm \chi} _{\left( {m,{p_m}} \right)}^{\left( k \right)} = \left\{ {{\bm P}_{\left( {m,{p_m}} \right)}^{\left( k \right)},{\bm E}_{\left( {m,{p_m}} \right)}^{\left( k \right)}} \right\}$.
The reflecting point ${\bm P}_{\left( {m,{p_m}} \right)}^{\left( k \right)}$ is the intersection point between the vector $\overrightarrow {{\bm r}_n^{\left( k \right)}{\bm r}_m^{\left( k \right)}} $ and the plane about which ${\bm x}_{bs}$ is symmetric with ${{\bm r}_n^{\left( k \right)}}$ as shown in Fig. \ref{fig_reflecting_element_edge_estimation}.
The normal vector of the reflecting surface is described by azimuth angle and polar angle which is denoted as $    {\bm E}_{\left( {m,{p_m}} \right)}^{\left( k \right)} = \left\{ {\bar \theta _{\left( {m,{p_m}} \right)}^{\left( k \right)},\bar \varphi _{\left( {m,{p_m}} \right)}^{\left( k \right)}} \right\} = {{\bm \kappa} ^{ - 1}}\left( {{{\bm x}_{bs}} - {\bm r}_n^{\left( k \right)}} \right)$,
where $n = \mathord{\buildrel{\lower3pt\hbox{$\scriptscriptstyle\frown$}}
\over o} _{\left( {m,{p_m}} \right)}^{\left( k \right)}$ is the CVT (index) associated with multi-path ${\left( {m,{p_m}} \right)}$ calculated by equation (\ref{eq_CODA_choose}), and ${{\bm \kappa} ^{ - 1}}\left(  \cdot  \right)$ is the inverse operation of ${{\bm \kappa} }\left(  \cdot  \right)$ that transforms the parameters of cartesian coordinate system to polar system.
The reflecting elements extracted based on CVT ${\bm C}_n^{\left(k\right)}$ (with $\mathord{\buildrel{\lower3pt\hbox{$\scriptscriptstyle\frown$}}
\over o} _{\left( {m,{p_m}} \right)}^{\left( k \right)}=n$) will be then collected together to  estimate the plane coordinate of the reflector $l_n$, where $l_n = \mathord{\buildrel{\lower3pt\hbox{$\scriptscriptstyle\frown$}}
\over \rho } _n^{\left( k \right)}$ is the reflector (index) associated with CVT ${\bm C}_n^{\left(k\right)}$ calculated by equation (\ref{eq_RCDA_choose}).

Then we can obtain the data set for the estimation of reflector ${\cal {\bm R}}_l$ as $\left\{ {{\bm \chi} _l^{\left( h \right)}} \right\} = \left\{ {{\bm P}_l^{\left( h \right)},{\bm E}_l^{\left( h \right)}} \right\} _{h = 1}^{{\cal H}_l}$, where ${\cal H}_l$ is the size of the set. ${{\bm \chi} _l^{\left( h \right)}}$ is the $h$-th reflecting element, ${\bm P}_l^{\left( h \right)}$ and ${\bm E}_l^{\left( h \right)} = \left\{ {{{\bar \theta }^{\left( h \right)}},{{\bar \varphi }^{\left( h \right)}}} \right\}$ are its reflecting point and the normal vector of the reflecting surface, respectively.

\subsection{FTRL based Reflector Learning and Edge Estimation}
\label{sec_EA_FTRL_EE}
\subsubsection{FTRL Reflector Learning}
For each reflector (e.g. ${\cal {\bm R}}_l$), an online learning architecture based on FTRL \cite{duchi2011adaptive} is introduced to estimate its plane coordinate with the data set $\left\{ {{\bm \chi} _l^{\left( h \right)}} \right\}$.
The estimating parameter is defined as ${\bm w} = \left( {\theta ,\varphi ,d} \right)^T$, which is also the basic parameter of a reflector as defined in Section \ref{sec_reflector_model}. The loss function is then defined as:
\begin{equation}\setlength{\abovedisplayskip}{3pt}\setlength{\belowdisplayskip}{3pt}
  J\left( w \right) = \frac{1}{{2{\cal H}}}\sum\limits_{h = 1}^{\cal H} {{{\left( {{\bm \kappa} {{\left( {\theta ;\varphi ;1} \right)}^T} \cdot {{\bm P}^{\left( h \right)}} + d} \right)}^2} + {\lambda _{{\rm{ref}}}}\left( {{{\left( {\theta  - {{\bar \theta }^{\left( h \right)}}} \right)}^2} + {{\left( {\varphi  - {{\bar \varphi }^{\left( h \right)}}} \right)}^2}} \right)} \label{eq_RL_FTRL_1}
\end{equation}
where $\lambda_{\rm ref}$ is the constraint weight for angle loss.

According to \cite{duchi2011adaptive,mcmahan2013ad}, the estimated parameter ${\bm w}$ can be updated as:
\begin{equation}\setlength{\abovedisplayskip}{3pt}\setlength{\belowdisplayskip}{3pt}
  {{w}_{t + 1,i}} = {-\eta _{t,i}}{z_{t,i}},\ {z_{t + 1,i}} = {z_{t,i}} + {g_{t,i}} - {\sigma _{t,i}}{w_{t,i}},\label{eq_FTRL_UPD}
\end{equation}
where ${{w}_{t,i}}$ is the $i$-th element in ${\bm w_t}$ at $t$-th iteration, ${\eta _{t,i}}$ is the learning rate based on a per-coordinate learning rate strategy \cite{mcmahan2013ad} defined as  ${\eta _{t,i}} = {{{l_\alpha }} \mathord{\left/
{\vphantom {{{l_\alpha }} {\left( {{l_\beta } + \sqrt {\sum\limits_{s = 1}^t {g_{s,i}^2} } } \right)}}} \right.
\kern-\nulldelimiterspace} {\left( {{l_\beta } + \sqrt {\sum\limits_{s = 1}^t {g_{s,i}^2} } } \right)}}$, ${z_{t,i}} $ is defined as ${z_{t,i}} = {g_{1:t,i}} - \sum\limits_{s = 1}^t {{\sigma _{s,i}}{w_{s,i}}}$ with ${{\sigma _{s,i}}}$ calculated as ${\sigma _{s,i}} = \frac{1}{{{\eta _{s,i}}}} - \frac{1}{{{\eta _{s - 1,i}}}}$, and ${g_{t,i}}$ is the gradient of $w_i$ at $t$-th iteration.
Setting ${n_i} =  {\sum\limits_{s = 1}^t {g_{s,i}^2} } $ and ${z_0} = 0$, the updating process in the $t$-th iteration can be summarized as:

{\em {\textbf {Step 1:}}} Calculate the parameter as ${{w}_{t + 1,i}} = {-\eta _{t,i}}{z_{t,i}}$ and the loss function as (\ref{eq_RL_FTRL_1}).

{\em {\textbf {Step 2:}}} For each $w_{t,i} \in {\bm w}_t$, i) calculate the gradient as ${g_{t,i}} = \nabla_{i} {J}\left( {{{\bm w}_t}} \right)$, ii) update the learning rate as ${\sigma _{t,i}} = \frac{{\sqrt {{n_i} + g_{t,i}^2}  - \sqrt {{n_i}} }}{\alpha }$, iii) update ${z_{t + 1,i}}$ as ${z_{t + 1,i}} = {z_{t,i}} + {g_{t,i}} - {\sigma _{t,i}}{w_{t,i}}$, iv) update $n_i$ as ${n_i} \leftarrow {n_i} + g_{t,i}^2$.

Based on the trained parameter $\mathord{\buildrel{\lower3pt\hbox{$\scriptscriptstyle\frown$}}
\over {\bm w}}  = {\left( {\mathord{\buildrel{\lower3pt\hbox{$\scriptscriptstyle\frown$}}
\over \theta } ,\mathord{\buildrel{\lower3pt\hbox{$\scriptscriptstyle\frown$}}
\over \varphi } ,\mathord{\buildrel{\lower3pt\hbox{$\scriptscriptstyle\frown$}}
\over d} } \right)^T}$ obtained from the above steps, the 3-D coordinate equation of a reflector (e.g. reflector ${\cal {\bm R}}_l$) can then be obtained as $\sin \mathord{\buildrel{\lower3pt\hbox{$\scriptscriptstyle\frown$}}
\over \varphi } \cos \mathord{\buildrel{\lower3pt\hbox{$\scriptscriptstyle\frown$}}
\over \theta } \cdot x + \sin \mathord{\buildrel{\lower3pt\hbox{$\scriptscriptstyle\frown$}}
\over \varphi } \sin \mathord{\buildrel{\lower3pt\hbox{$\scriptscriptstyle\frown$}}
\over \theta } \cdot y + \cos \mathord{\buildrel{\lower3pt\hbox{$\scriptscriptstyle\frown$}}
\over \varphi } \cdot z + \mathord{\buildrel{\lower3pt\hbox{$\scriptscriptstyle\frown$}}
\over d} = 0$.
Then the basic parameter of the reflector can be updated as ${{\bm w}_l} = \mathord{\buildrel{\lower3pt\hbox{$\scriptscriptstyle\frown$}}
\over {\bm w}}$, the normal vector can be updated as ${{\bm n}_l} = {\left( {\sin \mathord{\buildrel{\lower3pt\hbox{$\scriptscriptstyle\frown$}}
\over \varphi } \cos \mathord{\buildrel{\lower3pt\hbox{$\scriptscriptstyle\frown$}}
\over \theta } ,\sin \mathord{\buildrel{\lower3pt\hbox{$\scriptscriptstyle\frown$}}
\over \varphi } \sin \mathord{\buildrel{\lower3pt\hbox{$\scriptscriptstyle\frown$}}
\over \theta } ,\cos \mathord{\buildrel{\lower3pt\hbox{$\scriptscriptstyle\frown$}}
\over \varphi } } \right)^T}$,
and ${\bm R}_l$ symmetric with the base station about that reflector can then be calculated by (\ref{eq_EA_FTRL_R_l_cal}).

\subsubsection{Edge Estimation}

The edge of the reflecting surfaces will help to calculate the reflective probability in Section \ref{sec_RP_cal}, which is estimated in this subsection based on its corresponding reflecting elements.

For each reflecting element ${\bm \chi} _l^{\left( h \right)}$ of the reflector $\mathcal {\bm R}_l$, its projection points to the reflecting surface can be calculated as $\widehat {\bm P}_l^{\left( h \right)} = \left( {1 - {{\bm n}_l} \otimes {\bm n}_l^T} \right) \cdot {\bm P}_l^{\left( h \right)} - {d_l}{{\bm n}_l}$.
 We define the weight of each reflecting element for describing the reflecting surface's edge as:
\begin{equation}\setlength{\abovedisplayskip}{3pt}\setlength{\belowdisplayskip}{3pt}
  \begin{gathered}
    {J}\left( {\widehat {\bm P}_l^{\left( h \right)}} \right) = {\left( {{\bm \kappa} {{\left( {{\theta _l};{\varphi _l};1} \right)}^T} \cdot \widehat {\bm P}_l^{\left( h \right)} + {d_l}} \right)^2}
     + \lambda_{\rm ref} \left( {{{\left( {\bar \theta _l^{\left( h \right)} - {\theta _l}} \right)}^2} + {{\left( {\bar \varphi _l^{\left( h \right)} - {\varphi _l}} \right)}^2}} \right) \\
  \end{gathered}. \label{eq_EA_BE_2}
\end{equation}
\quad For reflector ${\cal R}_l$, an edge point ${\Omega _\theta ^l }$ can be estimated in each direction $\theta  \in \left[ {0,\left. {2\pi } \right)} \right.$ along with the ray starting from the center point ${{\widehat {\bm P}}_l} = {\mathbb E}\left( {\widehat {\bm P}_l^{\left( h \right)}} \right)$.
If ${\cal V}_l^\theta $ angles are sampled from $\left[ {0,\left. {2\pi } \right)} \right.$ represented as $\left\{ {{\theta _{\rm{1}}},...,{\theta _v},...,{\theta _{{\cal V}_l^\theta }}} \right\}$, the estimated edge can then be described as ${{\bm \Gamma}_l} = \left\{ {\Omega _{{\theta _1}}^l,...,\Omega _{{\theta _{{\cal V}_l^\theta }}}^l} \right\}$.
The point $\Omega _{{\theta _v}}^l$ is the edge estimated in $\theta_v$ direction, which is selected if the areal density of the weight calculated by equation (\ref{eq_EA_BE_2}) on the area from that point away from the central point $\widehat {\bm P}_l$ along with the directions ${{\bm \vartheta} _v}\left( {\nu _l^\theta } \right) \in \left[ {\left. {{\theta _v} - \frac{\pi }{{\nu _l^\theta }},{\theta _v} + \frac{\pi }{{\nu _l^\theta }}} \right)} \right.$ is just larger than the average areal density $ {{\bar W}_\theta }$:
\begin{equation}\setlength{\abovedisplayskip}{3pt}\setlength{\belowdisplayskip}{3pt}
  \Omega _{{\theta _v}}^l = \mathop {\arg }\limits_\Omega  \frac{{\int_{S_{{\theta _v}}^\Omega } {J\left( {\widehat {\bm P}_l^{\left( h \right)}} \right){\rm{d}}S} }}{{\int_{S_{{\theta _v}}^\Omega } {{\rm{d}}S} }} = {{\bar W}_\theta },\label{eq_EA_BE_4}
\end{equation}
where ${S_{\theta_v} ^\Omega }$ is the area of the surface integration in (\ref{eq_EA_BE_4}) shown in Fig. \ref{fig_reflecting_element_edge_estimation}. The average weight $ {{\bar W}_\theta }$ is calculated as ${{\bar W}_\theta } = \frac{{\int_{{S_{{\theta _{\left\{ v \right\}}}}}} {J\left( {\widehat {\bm P}_l^{\left( h \right)}} \right){\rm{d}}S} }}{{\int_{{S_{{\theta _{\left\{ v \right\}}}}}} {{\rm{d}}S} }}$, where ${{S_{\theta_v} }}$ is the area in ${{\bm \vartheta} _v}\left( {\nu _l^\theta } \right)$ directions starting from the central point $\widehat {\bm P}_l$ as shown in Fig. \ref{fig_reflecting_element_edge_estimation}.

\subsubsection{Reflective Probability}
\label{sec_RP_cal}
The the reflective probability ${p_{\mathbbm 1}}\left( {{\mathbbm 1}\left| {\bm x}_m \right.;{\mathcal{\bm R}_l}} \right)$ is calculated based on the ray-crossing algorithm \cite{sutherland1974characterization}.
According to the theory of ray-crossing, if a ray starting from a certain point have an odd number of intersection point(s) across the edge of the area, then that point is in the area enclosed by that edge. Thus for a vehicle locating at ${\bm r}_m$, we can first obtain its potential reflecting point  ${\bm P}_{ml}$ with respect to reflector ${\cal {\bm R}}_l$ (the intersection point between the vector $\overrightarrow {{{\bm r}_m}{{\bm R}_l}} $ and the plane coordinate of reflector ${\cal {\bm R}}_l$), and then the intersection points ${{\bm \Omega} _{ml}}$ between the ray from ${{ {\bm P}}_{ml}}$ along $-x$ direction and  the $l$-th reflector's edge ${\bm \Gamma}_l$ can be easily calculated.

  We use a binary random variable $\mathbbm 1_{m,l}$ to denote whether a vehicle locates at ${\bm r}_m$ can receive the signal reflected by the reflector ${\cal {\bm R}}_l$, and its probability according to the results of reflective sensing and mapping is denoted as:
\begin{equation}\setlength{\abovedisplayskip}{3pt}\setlength{\belowdisplayskip}{3pt}
  p\left( {{\mathbbm 1}_{m,l}=1\left| {{\bm r}_m,{\mathcal{\bm R}_l}} \right.} \right) = \left\{ \begin{array}{l}
  1,\,\bmod \,\left( {\left| {{{\bm \Omega} _{ml}}} \right|,2} \right) = 1\\
  0,otherwise
  \end{array} \right.. \label{eq_EA_BE_11}
\end{equation}

Since the reflective sensing and mapping continuously extracts information from cooperative SLAM for reflector estimation, we utilize the reflector density factor ${f_D}\left( {{{\cal {\bm R}}_l}} \right)$ to indicate the reliability of a reflector estimation. The reflector density of a reflecting surface is zero if the number of its reflecting elements ${\cal H}_l$ is smaller than the scaling number ${\cal H}_{\rm scale}$, in this case the estimation of that reflector is regarded unreliable.  If ${\cal H}_l > {\cal H}_{\rm scale}$, the reflecting density is defined as ${f_D}\left( {{\mathcal{R}_l}} \right) \triangleq 1 - {e^{ - {{{\mathcal{H}_l}} \mathord{\left/
{\vphantom {{{\mathcal{U}_l}} {\left( {{\mathcal{H}_{{\text{scale}}}}\int_{{\Gamma _l}} {{\text{d}}S} } \right)}}} \right.
\kern-\nulldelimiterspace} {\left( {{\mathcal{H}_{{\text{scale}}}}\int_{{\Gamma _l}} {{\text{d}}S} } \right)}}}}$.
Since the reflective probability is independent of velocity and synchronization bias, $p\left( {{\mathbbm 1}_{m,l}\left| {{\bm r}_m,{\mathcal{\bm R}_l}} \right.} \right)=p\left( {{\mathbbm 1}_{m,l}\left| {{\bm x}_m,{\mathcal{\bm R}_l}} \right.} \right)$.
Thus the reflective probability when ${\mathbbm 1}_{m,l}=1$ is defined as:
\begin{equation}
  {p_R}\left( {{\mathbbm 1}_{m,l}=1\left| {\bm x}_m \right.;{\mathcal{\bm R}_l}} \right) = {p\left( {{\mathbbm 1}_{m,l}=1\left| {{\bm x}_m,{\mathcal{\bm R}_l}} \right.} \right){f_D}\left( {{\mathcal{\bm R}_l}} \right) + f_0^R\left( {{\mathbbm 1}_{m,l}=1\left| {{\bm x}_m,{\mathcal{\bm R}_l}} \right.} \right)\left[ {1 - {f_D}\left( {{\mathcal{\bm R}_l}} \right)} \right]} ,\label{eq_reflective_prob_defind_1}
\end{equation}
where $f_0^R\left( {{\mathbbm 1}_{m,l}\left| {{\bm x}_m,{\mathcal{\bm R}_l}} \right.} \right)$ is the activation reflective probability in case that the reflector estimation is not reliable. 

\section{Radio Geometrization}
\label{sec_radio_geometrization}
This section introduces the radio geometrization to estimate the geometric paths of the multi-path propagation rays from the base station to a newcomer vehicle so as to further estimate its position and time synchronization bias. To this end, the radio geometrization component firstly introduces a Viterbi \cite{forney1973viterbi} based reflector decoding algorithm to link the multi-path observations from the newcomer vehicle to the  current reflectors, so that the reflectors that reflect the signal of the observed multi-paths can then be decoded. Then the radio geometrization component estimates the position and time synchronization bias of the vehicle based on the Separate Constrained Weighted Least Square (SCWLS) algorithm \cite{cheung2006constrained,Yan2016Improved}
 which is referred to as wake-up positioning and synchronization. Note that the radio geometrization can be easily extended from the downlink scenario in this section to an uplink scenario.

\subsection{Viterbi Based Reflector Decoding}
\label{sec_RG_Decoding}

Since the position of the newcomer vehicle is unknown, its multi-path observations cannot be modeled into VTs according to equation (\ref{eq_SM_5}). However, the relative positions among VTs can be utilized to decode a certain reflector for each multi-path observation, where that reflector has the biggest probability to be the one reflects the signals from the base station to the newcomer vehicle. We assume that there is no ternary isomorphism in the spatial distribution among all CVTs, and there are more than three multi-path components observed by the newcomer vehicle. The state space, transition matrix, observation space, emission matrix and initial probability of the Viterbi based reflector decoding process are then described as follows:

\subsubsection{State Space}

Based on the no ternary isomorphism assumption, we group three CVTs as a state. So the state space of the Viterbi based reflector decoding is defined as $ {\bm Q} = \left\{ {{{\bm q}_1},...,{{\bm q}_u},...,{{\bm q}_U}} \right\}$. The state ${\bm q}_u$ is defined as ${{\bm q}_u} = \left\{ {{\bm R}_{{l_1}}^T,{\bm R}_{{l_2}}^T,{\bm R}_{{l_3}}^T} \right\}$, where $\left( {{l_1},{l_2},{l_3}} \right)$ is a combinatorial number with $U = C_L^3$ kinds of combinations
\subsubsection{Observation Space}
The ToA and AoA observations of the newcomer vehicle are defined as ${{\bm Z}_x} = \left\{ {{{\bm z}_{\left( {x,1} \right)}},{{\bm z}_{\left( {x,2} \right)}}},\right.$ $\left.{ ...,{{\bm z}_{\left( {x,{P_x}} \right)}}} \right\}$,
where ${{{\bm z}_{\left( {x,p} \right)}}}$ refers to the observations of the $p$-th multi-path defined as ${{\bm z}_{\left( {x,p} \right)}} = \left\{ {{\theta _p},{\varphi _p},{d_p}} \right\}$, and $P_x$ is the number of multi-path observations. The observation is divided into $D = {P_x} - 2$ layers with three multi-path observations each. Then the observation of the $d$-th layer is defined as ($d=1,2,...,P_x-2$):
\begin{equation}\setlength{\abovedisplayskip}{3pt}\setlength{\belowdisplayskip}{3pt}
  {{\bm V}^{\left( d \right)}} = \left\{ {{{\bm z}_{\left( {x,d} \right)}^T},{{\bm z}_{\left( {x,d + 1} \right)}^T},{{\bm z}_{\left( {x,d + 2} \right)}^T}} \right\}.\label{eq_viterbi_ob_d_layer_def}
\end{equation}
\subsubsection{Emission Matrix}
The observation emission matrix of Viterbi based reflector decoding is defined as:
\begin{equation}\setlength{\abovedisplayskip}{3pt}\setlength{\belowdisplayskip}{3pt}
  \begin{array}{c}
  {\bm B}\left( {{{\bm V}^{\left( d \right)}}} \right) = {\left[ {{b_v}\left( {{{\bm V}^{\left( d \right)}}} \right)} \right]_{U \times 1}},\
  {b_v}\left( {{{\bm V}^{\left( d \right)}}} \right) = p\left( {{{\bm V}^{\left( d \right)}}\left| {{{\bm u}_d} = {{\bm q}_v}} \right.} \right)
  \end{array},\label{eq_viterbi_emission_def_m1}
\end{equation}
where ${b_v}\left( {{{\bm V}^{\left( d \right)}}} \right)$ is the probability of observing ${{{\bm V}^{\left( d \right)}}}$ when the state is ${\bm q}_v$.
The relative positions among the three VTs recast from the observations of $d$-th layer ${\bm V}^{\left( d \right)}$ are utilized to measure the probability of observing ${\bm V}^{\left( d \right)}$ when the state is ${\bm q}_v$.
Then ${b_v}\left( {{{\bm V}^{\left( d \right)}}} \right)$ is calculated as:
\begin{equation}\setlength{\abovedisplayskip}{3pt}\setlength{\belowdisplayskip}{3pt}
  \begin{gathered}
    {b_v}\left( {{{\bm V}^{\left( d \right)}}} \right) = g\left( {\varpi _v^{\left( d \right)};0,{\sigma ^2}} \right) \hfill ,\
    \varpi _v^{\left( d \right)} = \mathop {\min }\limits_{{{\bm E}_3^{\left( d \right)}}} \left\{ {{{\left\| {{{\bm q}_v}{{\bm E}_3^{\left( d \right)}}{\bm \Delta}  + {\bm \kappa} \left( {{{\bm V}^{\left( d \right)}}} \right){\bm \Delta} } \right\|}}} \right\} \hfill \\
  \end{gathered}, \label{eq_viterbi_mission_b_j_def_m1}
\end{equation}
where $g\left( {\varpi _v^{\left( d \right)};0,{\sigma ^2}} \right)$ denotes that ${\varpi _v^{\left( d \right)}}$ follows a zero mean Gaussian distribution with variance of $\sigma^2$, ${{{\bm E}_3^{\left( d \right)}}}$ is a $3 \times 3$ permutation matrix\footnotemark[3],
\footnotetext[3]{A permutation matrix is a square binary matrix that has exactly one entry of 1 in each row and each column, and 0s elsewhere. Specifically, for the dimension of $3 \times 3$, there are $A_3^3 = 6$ kinds of permutation matrixes.}
and ${\bm \Delta} $ is a subtraction matrix aiming to calculate relative position defined as {\small ${\bm \Delta}  = \left[  {1, - 1,0; - 1,0,1;0,1, - 1} \right]$}.
Specially, if ${\bm z}$ is a set of observations like ${\bm z} = \left( {{{\bm z}_1},{{\bm z}_2},...,{{\bm z}_N}} \right)$, then ${\bm \kappa} \left( {\bm z} \right) = \left( {{\bm \kappa} \left( {{{\bm z}_1}} \right),{\bm \kappa} \left( {{{\bm z}_2}} \right),...,{\bm \kappa} \left( {{{\bm z}_N}} \right)} \right)$.

\subsubsection{Transition Matrix}
The transition matrix of Viterbi based reflector decoding is defined as:
\begin{equation}\setlength{\abovedisplayskip}{3pt}\setlength{\belowdisplayskip}{3pt}
{\bm A}^{\left( d \right)} = {\left[ {{{a_{uv}^{\left( d \right)}}}} \right]_{U \times U}},{{a_{uv}^{\left( d \right)}}} = p\left( {{{\bm u}_{d + 1}} = {{\bm q}_v}\left| {{{\bm u}_d} = {{\bm q}_u}} \right.} \right),\label{eq_viterbi_transition_matrix_def}
\end{equation}
where ${\bm u}_d$ denotes the current state in $d$-th layer, and the value $a_{uv}^{\left( d \right)}$ means the probability that the state ${\bm q}_u$ transits to state ${\bm q}_v$ from $d$-th layer to $\left( {d + 1} \right)$-th layer. The value $a_{uv}^{\left( d \right)}$ is defined as:
\begin{equation}\setlength{\abovedisplayskip}{3pt}\setlength{\belowdisplayskip}{3pt}
  {a_{uv}^{\left( d \right)}} \triangleq \left\{ \begin{gathered}
  1,\left( {{{\bm q}_u}{\bm E}_3^{\left( d \right)}} \right) \oplus \left( {{{\bm q}_v}{\bm E}_3^{\left( {d + 1} \right)} >  > 1} \right) = \left( {\begin{array}{*{20}{c}}
  1&1&0
\end{array}} \right) \hfill \\
  0,otherwise \hfill \\
\end{gathered}  \right..\label{eq_viterbi_auv_def}
\end{equation}

The transition matrix in (\ref{eq_viterbi_auv_def})  indicates that the two states ${\bm q}_u$ and ${\bm q}_v$ in the transition process have only two same CVTs in specified place, which will ensure the relativity in the transition process according to the observation space defined in (\ref{eq_viterbi_ob_d_layer_def}).

\subsubsection{Initial Probability}
\label{sec_RG_init_P}

The initial probability of Viterbi based reflector decoding is defined as equation (\ref{eq_viterbi_initial_probability_def}), which constrains that there is no same CVT in any state.
\begin{equation}\setlength{\abovedisplayskip}{3pt}\setlength{\belowdisplayskip}{3pt}
{\bm \Pi}  = \left( {{\pi _1},{\pi _2},...,{\pi _U}} \right),{\pi _u} \propto \left\{ \begin{gathered}
    1,rank\left( {{{\bm q}_u}} \right) = 3 \hfill \\
    0,otherwise \hfill \\
  \end{gathered}  \right.. \label{eq_viterbi_initial_probability_def}
\end{equation}

The highest probability transiting to state ${\bm u}_d$ among all the possible state transition paths $\left( {{{\bm u}_1},{{\bm u}_2},...,{{\bm u}_{d - 1}}} \right)$ based on the observations ${\bm V}^{\left( {1:d} \right)}$ up to $d$-th layer  is defined as ${{\bm \delta} ^{\left( d \right)}}\left( {\bm u} \right)$,
\begin{equation}\setlength{\abovedisplayskip}{3pt}\setlength{\belowdisplayskip}{3pt}
  \begin{array}{l}
{{\bm \delta} ^{\left( d \right)}}\left( {\bm u} \right)
 =\mathop {\max }\limits_{{{\bm u}_1},{{\bm u}_2},...,{{\bm u}_{d - 1}}} p\left( {{{\bm u}_d} = {\bm u},{{\bm u}_1},{{\bm u}_2},...,{{\bm u}_{d - 1}},{{\bm V}^{\left( {1:d} \right)}}\left| {\bm \lambda}  \right.} \right) \end{array}.\label{eq_viterbi_lamda_def}
\end{equation}
\quad So its recursion formula of ${{\bm \delta} ^{\left( d \right)}}\left( {\bm u} \right)$ is derived as:
\begin{equation}\setlength{\abovedisplayskip}{3pt}\setlength{\belowdisplayskip}{3pt}
  \begin{gathered}
  {{\bm \delta} ^{\left( {d + 1} \right)}}\left( {\bm u} \right)  \hfill
   = \mathop {\max }\limits_{{{\bm u}_1},{{\bm u}_2},...,{{\bm u}_d}} p\left( {{{\bm u}_{d + 1}} = {\bm u},{{\bm u}_1},{{\bm u}_2},...,{{\bm u}_d},{V^{\left( {1:d + 1} \right)}}\left| {\bm \lambda}  \right.} \right) \hfill
   = \mathop {\max }\limits_{{\bm v} \in {\bm Q}} \left[ {{\delta ^{\left( d \right)}}\left( {\bm v} \right){a_{uv}^{\left( d \right)}}} \right]{b_v}\left( {{{\bm V}^{\left( {d + 1} \right)}}} \right).
\end{gathered} \label{eq_viterbi_lamda_recursion_def}
\end{equation}
\quad Thus the most probable state in ${{\left( {d - 1} \right)}}$-th layer transiting to the state ${\bm u}_d$ in $d$-th layer is defined as ${{\bm \Psi} ^{\left( d \right)}}\left( {\bm u} \right)$, which can be derived based on ${{\bm \delta} ^{\left( d \right)}}\left( {\bm u} \right)$ as:
\begin{equation}\setlength{\abovedisplayskip}{3pt}\setlength{\belowdisplayskip}{3pt}
  {{\bm \Psi} ^{\left( d \right)}}\left( {\bm u} \right) = \mathop {\arg \max }\limits_{{\bm v} \in {\bm Q}} \left[ {{{\bm \delta} ^{\left( {d - 1} \right)}}\left( {\bm v} \right){a_{uv}^{\left( d \right)}}} \right].\label{eq_viterbi_FAI_def}
\end{equation}
\quad Then the algorithm for Viterbi based reflector decoding can be summarized as:

{\em {\textbf {Step 1:}}} Initialize ${{\bm \delta} ^{\left( 1 \right)}}\left( {\bm u} \right)$ and $  {\Psi ^{\left( {1} \right)}}\left( {\bm u} \right)$ as  ${{\bm \delta} ^{\left( 1 \right)}}\left( {\bm u} \right) = {\pi _u}{b_u}\left( {{{\bm V}^{\left( 1 \right)}}} \right),\
  {\Psi ^{\left( {1} \right)}}\left( {\bm u} \right) = 0,{\bm u} \in {\bm Q}$.

{\em {\textbf {Step 2:}}} Update the state in each layer dynamically as (\ref{eq_viterbi_lamda_recursion_def}) and (\ref{eq_viterbi_FAI_def}).

{\em {\textbf {Step 3:}}} Calculate the probability of the most probable path and its state in layer $D$ as ${P^*} = \mathop {\max }\limits_{{\bm u} \in {\bm Q}} {{\bm \delta} ^{\left( D \right)}}\left( {\bm u} \right),{\bm q}_D^* = \mathop {\arg \max }\limits_{{\bm u} \in {\bm Q}} \left[ {{{\bm \delta} ^{\left( D \right)}}\left( {\bm u} \right)} \right]$.

{\em {\textbf {Step 4:}}} Recall the states in layer $d = D - 1,...,1$ through $\Psi \left( {\bm u} \right)$ as
${\bm q}_d^* = {\Psi ^{\left( {d + 1} \right)}}\left( {{\bm q}_{d + 1}^*} \right)$.

Then the most probable state ${\bm q}_d^*$ corresponding to the observation ${{{\bm V}^{\left( d \right)}}}$ is decoded as ${{\bm V}^{\left( d \right)}} \sim {\bm q}_d^*{\bm E}_3^{\left( d \right)}$. Thus the most probable CVTs (generated by reflectors) $\left\{ {{{\bm R}_{\left( {x,p} \right)}}} \right\}_{p=1}^{P_x}$ corresponding to the observation ${\bm Z}_x$ are then decoded as $  {{\bm Z}_x} \sim  \left\{ {{{\bm R}_{\left( {x,1} \right)}},{{\bm R}_{\left( {x,2} \right)}},...,{{\bm R}_{\left( {x,{P_x}} \right)}}} \right\}$.
\subsection{Wake-up Positioning and Synchronization}
\label{sec_RG_SCWLS}

This section utilizes the reflector decoding results in Section \ref{sec_RG_Decoding} to estimate the position and time synchronization bias of the newcomer vehicle to achieve wake-up positioning and synchronization based on the SCWLS algorithm \cite{cheung2006constrained,Yan2016Improved}. The ToA observations $\left\{ {{d_p}} \right\}$  are utilized to achieve TDoA positioning, and the AoA observations $\left\{ {{\theta _p},{\varphi _p}} \right\}$ are utilized to achieve AoA positioning. Finally, the two positioning methods are jointly considered in the wake-up positioning and synchronization algorithm.

\subsubsection{TDoA Positioning}

Since there exists time synchronization bias, the ToA measurement $d_p$ can be resolved as
$  {d_p} = {\delta _p} + b + n_p^d\label{eq_scwls_1}$,
where ${\delta _p}$ is the real distance from CVT ${{\bm R}_{\left( {x,p} \right)}}$ to the vehicle $x$, $b$ is the time synchronization bias, and $n_p^d$ is the observation error of $d_p$.
We define the ToA observation of the first multi-path $d_1$ as reference ToA, so the distance difference between the ToA observation of multi-path $p$ and the reference ToA is calculated as:
\begin{equation}\setlength{\abovedisplayskip}{3pt}\setlength{\belowdisplayskip}{3pt}
{r_{p,1}} = {d_p} - {d_1} = {\delta _p} - {\delta _1} + n_{p,1}^d, \label{eq_scwls_2}
\end{equation}
where $p = 1,2,...,{P_x}$ denote the indexes of the multi-path,
 and ${n_{p,1}^d}$ is the error of the distance difference calculated as $n_{p,1}^d = n_p^d - n_p^d$.
For sake of simplification, we define ${{\bm R}_{\left( {x,p} \right)}} = {\left( {{x_p},{y_p},{z_p}} \right)^T}$, and the position of the newcomer vehicle as ${{\bm r}_x} = \left( {x,y,z} \right)^T$. Then $\delta_p$ is calculated as ${\delta _p} = {\left\| {{{\bm R}_{\left( {x,p} \right)}} - {{\bm r}_x}} \right\|_F}$.
According to (\ref{eq_scwls_2}), we can obtain:
\begin{equation}\setlength{\abovedisplayskip}{3pt}\setlength{\belowdisplayskip}{3pt}
  \begin{gathered}
    {\bm E}{\bm \vartheta}  = {\bm h} + {\bm m}  ,\
    {\bm \vartheta}  = {\left( {\begin{array}{*{20}{c}}
    {x - {x_1}}&{y - {y_1}}&{z - {z_1}}&{{\delta _1}}
  \end{array}} \right)^T},
  {\bm m}  = {\left( {\begin{array}{*{20}{c}}
  {{m _{2,1}}}& \cdots &{{m _{{P_x},1}}}
\end{array}} \right)^T}, \\
  {\bm E} = \left( {\begin{array}{*{20}{c}}
  {{x_2} - {x_1}}&{{y_2} - {y_1}}&{{z_2} - {z_1}}&{{r_{2,1}}} \\
  \vdots & \vdots & \vdots & \vdots  \\
  {{x_{{P_x}}} - {x_1}}&{{y_{{P_x}}} - {y_1}}&{{z_{{P_x}}} - {z_1}}&{{r_{{P_x},1}}}
  \end{array}} \right),
  {\bm h} = \frac{1}{2}\left( {\begin{array}{*{20}{c}}
    {\left\| {{{\bm R}_{\left( {x,2} \right)}} - {{\bm R}_{\left( {x,1} \right)}}} \right\|^2 - r_{2,1}^2}\\
     \vdots \\
    {\left\| {{{\bm R}_{\left( {x,{P_x}} \right)}} - {{\bm R}_{\left( {x,1} \right)}}} \right\|^2 - r_{{P_x},1}^2}
    \end{array}} \right),
  \end{gathered}  \label{eq_scwls_3}
\end{equation}
\noindent where ${m_{p,1}} = {\delta _p}n_{p,1}^d + {{{{\left( {n_{p,1}^d} \right)}^2}} \mathord{\left/
 {\vphantom {{{{\left( {n_{p,1}^d} \right)}^2}} 2}} \right.
 \kern-\nulldelimiterspace} 2}$. If the second-order error is ignored, then ${m_{p,1}}  \approx  {\delta _p}n_{p,1}^d$.

\subsubsection{AOA Positioning}

For the observation of azimuth angle $\theta _p$, we have ${\theta _p} = \theta _p^r + n_p^\theta $, where $\theta _p^r$ denotes the real azimuth angle and $n_p^\theta $ denotes its observation error. According to the definition of azimuth angle, we have $  {{\left( {y - {y_p}} \right)} \mathord{\left/
{\vphantom {{\left( {y - {y_p}} \right)} {\left( {x - {x_p}} \right)}}} \right.
\kern-\nulldelimiterspace} {\left( {x - {x_p}} \right)}} = {{\sin \theta _p^r} \mathord{\left/
{\vphantom {{\sin \theta _p^r} {\cos \theta _p^r}}} \right.
\kern-\nulldelimiterspace} {\cos \theta _p^r}}$.
For the observation of polar angle $\varphi _p$, we define ${\varphi _p} = \varphi _p^r + n_p^\varphi $, where $\varphi _p^r$ denotes the real polar angle, and $n_p^\varphi$ denotes its observation error. According to the definition of polar angle, we can obtain
 $  \cos \varphi _p^r = \frac{{z - {z_p}}}{{{\delta _p}}}$. So we can get the following equation for the AoA positioning:
 \begin{equation}\setlength{\abovedisplayskip}{3pt}\setlength{\belowdisplayskip}{3pt}
  \begin{gathered}
    {{\bm H}}{{\bm \vartheta}} = {{\bm K}} + {\bm \mu} + {\bm \nu}  ,\
  {\bm \mu}  = {\left( {\begin{array}{*{20}{c}}
  {{\mu _1}}& \cdots &{{\mu _{{P_x}}}}
\end{array}} \right)^T} ,\
{\bm \nu}  = {\left( {\begin{array}{*{20}{c}}
  {{\nu _1}}& \cdots &{{\nu _{{P_x}}}}
\end{array}} \right)^T},\\
  {\bm H} = \left( {\begin{array}{*{20}{c}}
  {\sin {\theta _1}}&{ - \cos {\theta _1}}&1&{ - \cos {\varphi _1}}\\
   \vdots & \vdots & \vdots & \vdots \\
  {\sin {\theta _{{P_x}}}}&{ - \cos {\theta _{{P_x}}}}&1&{ - \cos {\varphi _{{P_x}}}}
  \end{array}} \right) ,\
  {\bm K} = \left( {\begin{array}{*{20}{c}}
    0\\
    {{r_{2,1}}\cos {\varphi _2}}\\
     \vdots \\
    {{r_{{P_x},1}}\cos {\varphi _{{P_x}}}}
    \end{array}} \right),
  \end{gathered}  \label{eq_scwls_5}
\end{equation}
where ${\mu _p} = n_p^\theta \left[ {\left( {x - {x_p}} \right)\cos {\theta _p} + \left( {y - {y_p}} \right)\sin {\theta _p}} \right]
 \approx n_p^\theta {\delta _p}\sin {\varphi _p}$, and ${\nu _p} = n_p^\varphi {\delta _p}\sin {\varphi _p} - n_{p,1}^d\cos {\varphi _p}$.

\subsubsection{Joint TDoA and AoA Positioning}
Considering the equations (\ref{eq_scwls_3}) and (\ref{eq_scwls_5}) simultaneously, we can obtain:
\begin{equation}\setlength{\abovedisplayskip}{3pt}\setlength{\belowdisplayskip}{3pt}
  \begin{gathered}
  {\bm A}{\bm \vartheta}  = {\bm q} + {\bm N} ,\
  {\bm A} = \left[ {\begin{array}{*{20}{c}}
    {\bm E}^{\rm T}&{\bm H}^{\rm T}
    \end{array}} \right]^{\rm T} ,\
  {\bm q} = \left[ {\begin{array}{*{20}{c}}
    {\bm h}^{\rm T}&{\bm K}^{\rm T}
    \end{array}} \right]^{\rm T},
  {\bm N} = \left[ {\begin{array}{*{20}{c}}
    {\bm m}^{\rm T}&\left( {\bm \mu}  + {\bm \nu} \right)^{\rm T}
    \end{array}} \right]^{\rm T}
\end{gathered}. \label{eq_scwls_9}
\end{equation}

Since there are potential inverse operations for a singular matrix in (\ref{eq_scwls_9}), the equation (\ref{eq_scwls_9})  can be  transformed as:
{{\begin{equation}\setlength{\abovedisplayskip}{3pt}\setlength{\belowdisplayskip}{3pt}
  \begin{gathered}
  {\bm G}{\bm \chi}  = {\bm q} - {\bm g}{\delta _1} + {\bm N} ,
  {\bm G} = {{\left( {\begin{array}{*{3}{c}}
  {{{\left[ {\bm A} \right]}_{:,1}}}&{{{\left[ {\bm A} \right]}_{:,2}}}&{{{\left[ {\bm A} \right]}_{:,3}}}
\end{array}} \right)}} ,
  {\bm \chi}  = {{{\left( {\begin{array}{*{3}{c}}
  {x - {x_1}}&{y - {y_1}}&{z - {z_1}}
\end{array}} \right)^T}}} ,
  {\bm g} = {\left[ {\bm A} \right]_{:,4}}
\end{gathered}, \label{eq_scwls_10}
\end{equation}}}%
where ${\left[ {\bm A} \right]_{:,i}}$ is the $i$-th row of matrix ${\bm A}$, and the weight matrix is defined as ${\bm W} = {\mathbb E}\left( {{\bm N}{{\bm N}^T}} \right)$.

Since the observation error from distance, azimuth angle, and polar angle are independent, the weight matrix can then be calculated as (the value ${\delta _p}$ in the calculation process of ${\bm W}$ is approximated as ${\delta _p} \approx {r_{p,1}} + {{\mathord{\buildrel{\lower3pt\hbox{$\scriptscriptstyle\frown$}}
\over \delta } } _1}$):
\begin{equation}\setlength{\abovedisplayskip}{3pt}\setlength{\belowdisplayskip}{3pt}
  \begin{array}{*{20}{l}}
    {{\bm W} = {{{\left( {\begin{array}{*{20}{c}}
    {\cal {\bm M}}&{\cal {\bm E}} \\
    {{{\cal {\bm E}}^T}}&{\cal {\bm N}}
  \end{array}} \right)}^{ - 1}}}},
  {{\cal {\bm E}} = {\mathbb E}\left( {{\bm m}{{\left( {{\bm \mu}  + {\bm \nu} } \right)}^T}} \right)},
    {{\cal {\bm M}} = {\mathbb E}\left( {{\bm m}{{\bm m}^T}} \right) },
    {{\cal {\bm N}} = {\mathbb E}\left( {\left( {{\bm \mu}  + {\bm \nu} } \right){{\left( {{\bm \mu}  + {\bm \nu} } \right)}^T}} \right)}
  \end{array}.\label{eq_scwls_10_1}
\end{equation}
\quad We define the weighted least square function of equation (\ref{eq_scwls_10}) as
$
{J_{LS}}\left( {{\mathord{\buildrel{\lower3pt\hbox{$\scriptscriptstyle\frown$}}
\over {\bm \chi} } } ,{{{\mathord{\buildrel{\lower3pt\hbox{$\scriptscriptstyle\frown$}}
\over \delta } } }_1}} \right) = {\left( {{\bm G}{\mathord{\buildrel{\lower3pt\hbox{$\scriptscriptstyle\frown$}}
\over {\bm \chi} } }  - {\bm q} + {\bm g}{{{\mathord{\buildrel{\lower3pt\hbox{$\scriptscriptstyle\frown$}}
\over \delta } } }_1}} \right)^T}\allowbreak \cdot{\bm W}\cdot\left( {{\bm G}{\mathord{\buildrel{\lower3pt\hbox{$\scriptscriptstyle\frown$}}
\over {\bm \chi} } }  - {\bm q} + {\bm g}{{{\mathord{\buildrel{\lower3pt\hbox{$\scriptscriptstyle\frown$}}
\over \delta } } }_1}} \right)\label{eq_scwls_11}
$, and ${\bm \chi}$ can then be estimated as:
\begin{equation}\setlength{\abovedisplayskip}{3pt}\setlength{\belowdisplayskip}{3pt}
  \begin{gathered}
    {{\mathord{\buildrel{\lower3pt\hbox{$\scriptscriptstyle\frown$}}
    \over {\bm \chi} } }}  = \mathop {\arg \min }\limits_{\mathord{\buildrel{\lower3pt\hbox{$\scriptscriptstyle\frown$}}
    \over {\bm \chi} }  } {J_{LS}}\left( { {{\mathord{\buildrel{\lower3pt\hbox{$\scriptscriptstyle\frown$}}
    \over {\bm \chi} } }} ,{{{\mathord{\buildrel{\lower3pt\hbox{$\scriptscriptstyle\frown$}}
    \over \delta } } }_1}} \right) \hfill \ \
    {\text{s}}.{\text{t}}.\ \ \ {\mathord{\buildrel{\lower3pt\hbox{$\scriptscriptstyle\frown$}}
    \over {\bm \chi} } ^T}\mathord{\buildrel{\lower3pt\hbox{$\scriptscriptstyle\frown$}}
    \over {\bm \chi} }  = {\mathord{\buildrel{\lower3pt\hbox{$\scriptscriptstyle\frown$}}
    \over \delta } _1}^2 \hfill \\
  \end{gathered}.    \label{eq_scwls_12}
\end{equation}
\quad The problem in (\ref{eq_scwls_12}) is solved by introducing the Lagrange multipliers $\eta$, then ${{\mathord{\buildrel{\lower3pt\hbox{$\scriptscriptstyle\frown$}}
\over {\bm \chi} } } }$ can be calculated iteratively by the co-called SCWLS algorithm. Initializing  ${\bm W} = {{\bm I}_{{2P_x} - 1}}$, the SCWLS algorithm can be summarized as:

{\em {\textbf {Step 1:}}}  Calculate the Lagrange multipliers $\eta $ as \cite{Yan2016Improved}, so that $\eta $ can be obtained with multiple roots: ${\eta ^{[s]}},s = 1,2,..S$, $S \leqslant 6$.

{\em {\textbf {Step 2:}}}  Calculate ${{{\mathord{\buildrel{\lower3pt\hbox{$\scriptscriptstyle\frown$}}
\over \delta } } }_1}$ based on $\left\{ {{\eta ^{[s]}}} \right\}$, and find the group of $\left\{ {{\eta ^{[s]}},{\mathord{\buildrel{\lower3pt\hbox{$\scriptscriptstyle\frown$}}
\over \delta } } _1^{\left[ s \right]},{{{\mathord{\buildrel{\lower3pt\hbox{$\scriptscriptstyle\frown$}}
\over {\bm \chi} } } }^{\left[ s \right]}}} \right\}$ that minimizes the Lagrangian.

{\em {\textbf {Step 3:}}}  Reconstruct ${\bm W}$ as (\ref{eq_scwls_10_1}), and repeat steps (1$\sim$2) until ${{\mathord{\buildrel{\lower3pt\hbox{$\scriptscriptstyle\frown$}}
\over {\bm \chi} } } }$ converges.

Thus the position and time synchronization bias of the unknown vehicle is calculated by (\ref{eq_scwls_18}), which can provide accurate initial positioning and time synchronization for cooperative SLAM.
\begin{equation}\setlength{\abovedisplayskip}{3pt}\setlength{\belowdisplayskip}{3pt}
  \begin{array}{l}
{{{\mathord{\buildrel{\lower3pt\hbox{$\scriptscriptstyle\frown$}}
\over {\bm r}} }}_x} = {\mathord{\buildrel{\lower3pt\hbox{$\scriptscriptstyle\frown$}}
\over {\bm \chi} } } + {{\bm R}_{\left( {x,1} \right)}},\
{\mathord{\buildrel{\lower3pt\hbox{$\scriptscriptstyle\frown$}}
\over b} } = {d_1} - {\left\| {{\mathord{\buildrel{\lower3pt\hbox{$\scriptscriptstyle\frown$}}
\over {\bm \chi} } } } \right\|_F}
\end{array}.\label{eq_scwls_18}
\end{equation}

\begin{figure*}
  \setlength{\abovecaptionskip}{-0.2cm}
  \centering
  \includegraphics[width=0.8\columnwidth]{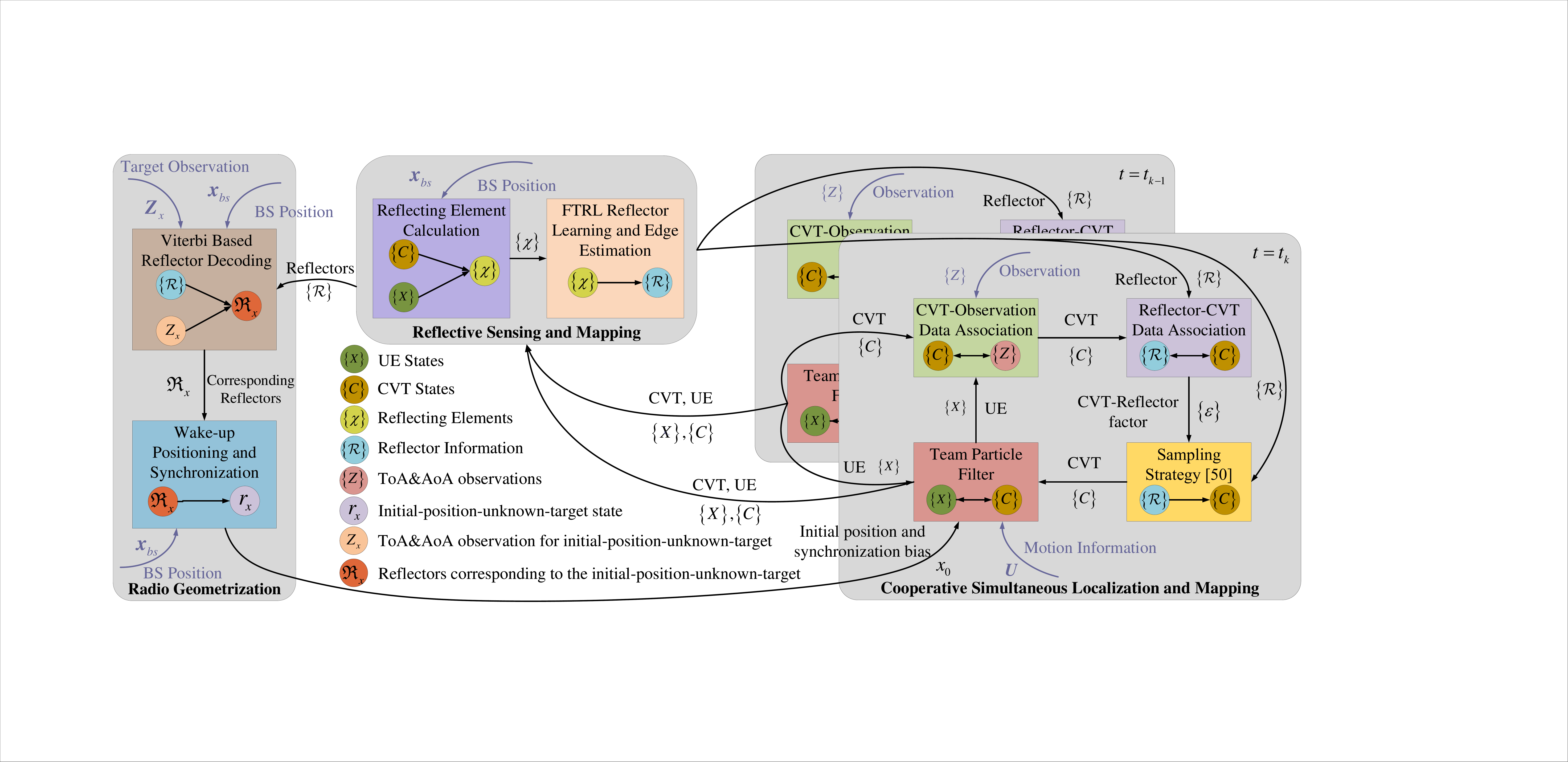}
  \caption{The three main components of TCSE are shown above: (i) cooperative SLAM,  (ii) Reflective sensing and mapping, (iii) Radio geometrization.
  }
  \label{TG}
  \vspace{-4ex}
\end{figure*}

\section{Overview of Implementation for Team Channel-SLAM Evolution}
\label{sec_implementation}

The relations among the three main algorithm components in TCSE are shown in Fig. \ref{TG} and is also summarized below for the reader's overall understanding.
More specifically, when TCSE is initially deployed, the cooperative SLAM component is initiated to estimate the state of CVTs and multiple vehicles jointly through the CVT-observation data association method and the team particle filter as described in Section \ref{sec_CO_SLAM}.
Meanwhile, the reflective sensing and mapping component collects the estimation results of multiple vehicles and CVTs from the cooperative SLAM to estimate the position and edge of the reflecting surface as described in Section \ref{sec_environmental_awareness}.

When the reflecting surfaces have been mapped up in sufficient detail, the framework can in turn improve the accuracy of cooperative SLAM by: 1) sampling the CVT particles based on reflector-CVT data association and the sampling strategy described in Section \ref{sec_CO_SLAM}, and 2) providing the reflective probability for CVT-observation data association and reflector-CVT data association as described in Section \ref{sec_environmental_awareness} and \ref{sec_CO_SLAM}.
Meanwhile, radio geometrization utilizes the information of reflecting surfaces to achieve wake-up positioning and synchronization as described in Section \ref{sec_radio_geometrization}.

\begin{figure*}[t]
  \setlength{\abovecaptionskip}{-0.2cm}
  \centering
  \includegraphics[width=0.8\columnwidth,height=7.2cm]{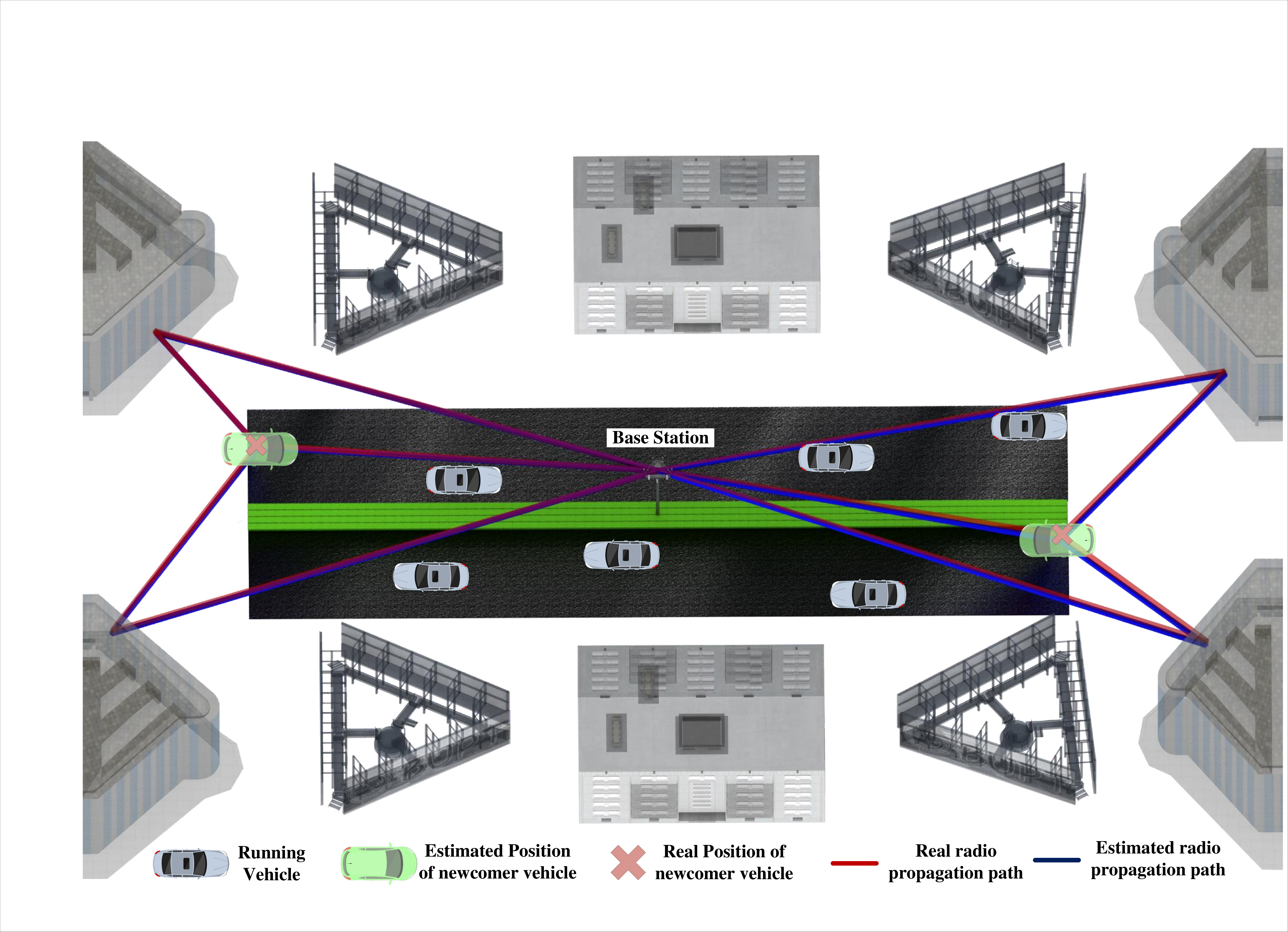}
  \caption{Simulation layout with road, running vehicles and surrounding reflecting buildings. The figure shows the accuracy of vehicle localization for two newcomer vehicle examples (wake-up positioning)  along with the reconstructed radio path for them (radio geometrization).
  }
  \label{fig_scene}
  \vspace{-4ex}
\end{figure*}

\section{Numerical Results}
\label{sec_numberical_results}
Simulations are done to test the performance of the TCSE algorithm. As shown in Fig. \ref{fig_scene}, the simulations in this paper focus on a fixed $100 {\rm m}$-length road with vehicles passing by with a certain flowing density, and a base station is placed with a known location at $\left[ {50{\rm{m}},0,8{\rm{m}}} \right]$.
{{
  There are 10 reflectors in our simulation as shown in Fig. \ref{fig_scene}, and the information of each reflectors (including the number of those reflecting surfaces) is unknown.
}}
The density of vehicles on the road is set to $8\ {\rm{vehicles}}/100\ {\rm{meters}}$, and the velocity of each vehicle is ${\rm{10m/s}}$. The length of each time slot ${t_\delta }=0.1{\rm s}$, and we execute TCSE for 50000 time slots in 100 simulation runs.
In order to verify the increased performance of
TCSE as the number of vehicles passing through the 100 meters’ road in increasing, we define the index of each vehicle passing through the road as $f_V$, which is referred to in the plots as {\em Vehicle Index}.  Hence ${f_V} = 1,2,...,f,...,F$, where $F$ is the total number of vehicles exposed in the experiments (in our case, $F = 3924$).

In order to be more realistic, we consider the possibility of measurement errors in the ToA, AoA, vehicle velocity, and initial localization.
The standard deviation of the ToA measurement error ($\sigma_d$) is set to $0.2{\rm m}$ following a zero-mean Gaussian distribution (ZMGD), and the standard deviation of AoA measurement error ($\sigma_{\theta}$, both for polar angle and azimuth angle) is set to 1 degree following a ZMGD. The magnitude error of velocity follows a ZMGD with standard deviation ${\sigma _v} = 0.1{\rm{m}}/{\rm{s}}$, and the orientation error of the velocity follows a ZMGD with standard deviation $\sigma _v^\theta  = 0.1{\rm{deg}}$.

The initial positioning is set with an error following a ZMGD with standard deviation ${\sigma _G} = 5{\text{m}}$. Note that this value is in line with typical GPS performance and corresponds to a large amount of initial uncertainty in relation to the objective of this work which is to achieve well sub-meter accuracy reliably (i.e. even when GPS signals have dropped). The time synchronization bias multiplied by speed of light follows a ZMGD with standard deviation ${\sigma _S} = 5{\rm{m}}$. The number of particles for vehicles and CVTs is set as ${\mathcal{N}_V} ={\mathcal{N}_C} = 120$.
{{
  The probability distribution of false alarm measurement ${{f_{{\text{FA}}}}\left( {z_{\left( {m,{p_m}} \right)}^{\left( k \right)}} \right)}$ is uniform on $\left[ {0,50{\rm{m}}} \right]$, $\left[ {0,2\pi } \right]$, and $\left[ {0,\pi } \right]$ for ToA, azimuth angle of AoA, and polar angle of AoA respectively. The further parameters mentioned in this paper are shown in Table \ref{tab:addlabel}.
}}

\renewcommand\arraystretch{1}
\begin{table}[t]
  \vspace{-0.5cm}
  \centering
  \setlength{\abovecaptionskip}{0pt}%
  \setlength{\belowcaptionskip}{0pt}%
  \caption{Parameter settings}
    \begin{tabular}{c|c|c|c|c|c|c|c|c|c}
    \Xhline{1pt}
    
    ${\mu _{{\text{FA}}}}$    & ${p_d}$     & $\delta_{\rm FA}$     & $\alpha_C$     & $p_M$     & $l_{\alpha}$     & $l_{\beta}$     & $\lambda_{\rm ref}$     & $f_0^R\left( {{\mathbbm 1}\left| {{\bm x}_m,{\mathcal{\bm R}_l}} \right.} \right)$     & ${\mathcal{H}_{{\text{scale}}}}$ \\
    \Xhline{0.6pt}
    $0, 0.5, 1$    & $1, 0.95, 0.9$    & ${10^{ - 4}}$    & $0.95$    & $0.05$    & $1.98\times10^{-3}$    & $0.99$    & $20 \times{\left( {{{{\sigma _d}} \mathord{\left/
    {\vphantom {{{\sigma _d}} {{\sigma _\theta }}}} \right.
    \kern-\nulldelimiterspace} {{\sigma _\theta }}}} \right)^2}$    & $0.5$    & $100$ \\
    \Xhline{1pt}
    \end{tabular}%
  \label{tab:addlabel}%
  \vspace{-4ex}
\end{table}%

We denote the vehicle positioning error and vehicle synchronization error (multiplied by speed of light) as ${\varepsilon _{VP}}$ and  ${\varepsilon _{VS}}$, respectively. Since the height of a vehicle is easy to access, we consider the 2-D positioning error for each vehicle. The vehicle positioning error after $f$ vehicle passed by is calculated as $\varepsilon _{VP}^{\left( f \right)} = \frac{1}{{{k_f} - {k_{f - 1}}}} \cdot \frac{1}{M} \cdot \sum\limits_{k = {k_{f - 1}} + 1}^{{k_f}} {\sum\limits_{m = 1}^M {{{\left\| { {\rm{ }}\mathord{\buildrel{\lower3pt\hbox{$\scriptscriptstyle\frown$}}
\over {\bm r}} _m^{\left( k \right)}-{\bm r}_{m,{\rm real}}^{\left( k \right)}} \right\|}_F}} } $, where $k_f$ is the time slot at which the $f$ vehicle is passing. ${\rm{ }}\mathord{\buildrel{\lower3pt\hbox{$\scriptscriptstyle\frown$}}
\over {\bm r}} _m^{\left( k \right)}$ and ${\bm r}_{m,{\rm real}}^{\left( k \right)}$ are the estimated and real positions of vehicle $m$ at time slot $t_k$. The time synchronization error of vehicles ${\varepsilon _{VS}}$ is calculated similarly.

\subsection{Reflective Sensing and Mapping}
\begin{figure*}[t]
  \setlength{\abovecaptionskip}{-0.2cm}
  \centering
  \includegraphics[width=0.83\columnwidth,height=5.6cm]{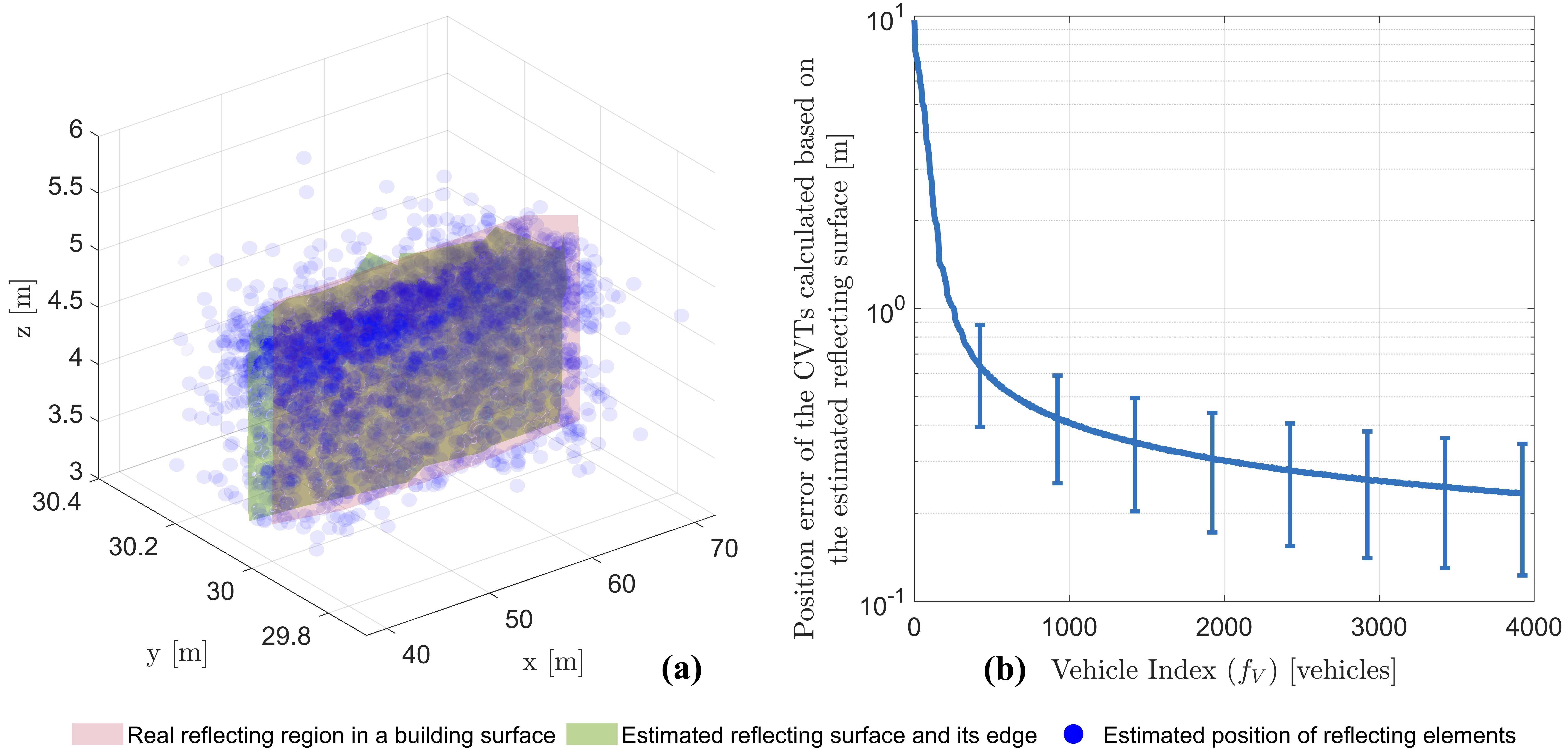}
  \caption{Performance of reflective sensing and mapping. In (a) the position and edge of a reflecting surface is reconstructed. In (b) it is seen how the common virtual transmitters (CVT) are better localized as the number of vehicles driving over time over the road section is increasing.
  }
  \label{fig_reflective_sensing_and_mapping_performance}
  \vspace{-2ex}
\end{figure*}

The reflective sensing and mapping component extracts the reflecting elements from the cooperative SLAM component to estimate the position and edge of the reflecting surfaces in an online learning way, and its performance is shown in Fig. \ref{fig_reflective_sensing_and_mapping_performance}. In Fig. \ref{fig_reflective_sensing_and_mapping_performance}(a), the blue colored balls indicate the reflecting elements, based on which the position of the reflecting surface and its edge are estimated (denoted by green surface). Then the CVT ${{\bm R}_{\left\{ l \right\}}}$ symmetric with the base station about the estimated surface can be calculated by equation (\ref{eq_EA_FTRL_R_l_cal}), and its estimation error over Vehicle Index can be seen in Fig. \ref{fig_reflective_sensing_and_mapping_performance}(b). We can see from the Fig. \ref{fig_reflective_sensing_and_mapping_performance}(a) that the position and edge of the reflecting surface are well close to the real reflecting surface, which means that the reflective sensing and mapping component can estimate the reflecting surface well based on the reflecting elements extracted from cooperative SLAM component. We can also see from Fig. \ref{fig_reflective_sensing_and_mapping_performance}(b) that the mean position error of the CVTs corresponding to the reflecting surfaces gets gradually decrease over the increasing of Vehicle Index (up to an error of $0.234{\rm m}$ until the last vehicle passed by), which further indicates that the reflective sensing and mapping component have a good performance in the reflecting surface estimation.

\begin{figure*}[t]
  \setlength{\abovecaptionskip}{-0.2cm}
  \centering
  \includegraphics[width=0.83\columnwidth,height=5.6cm]{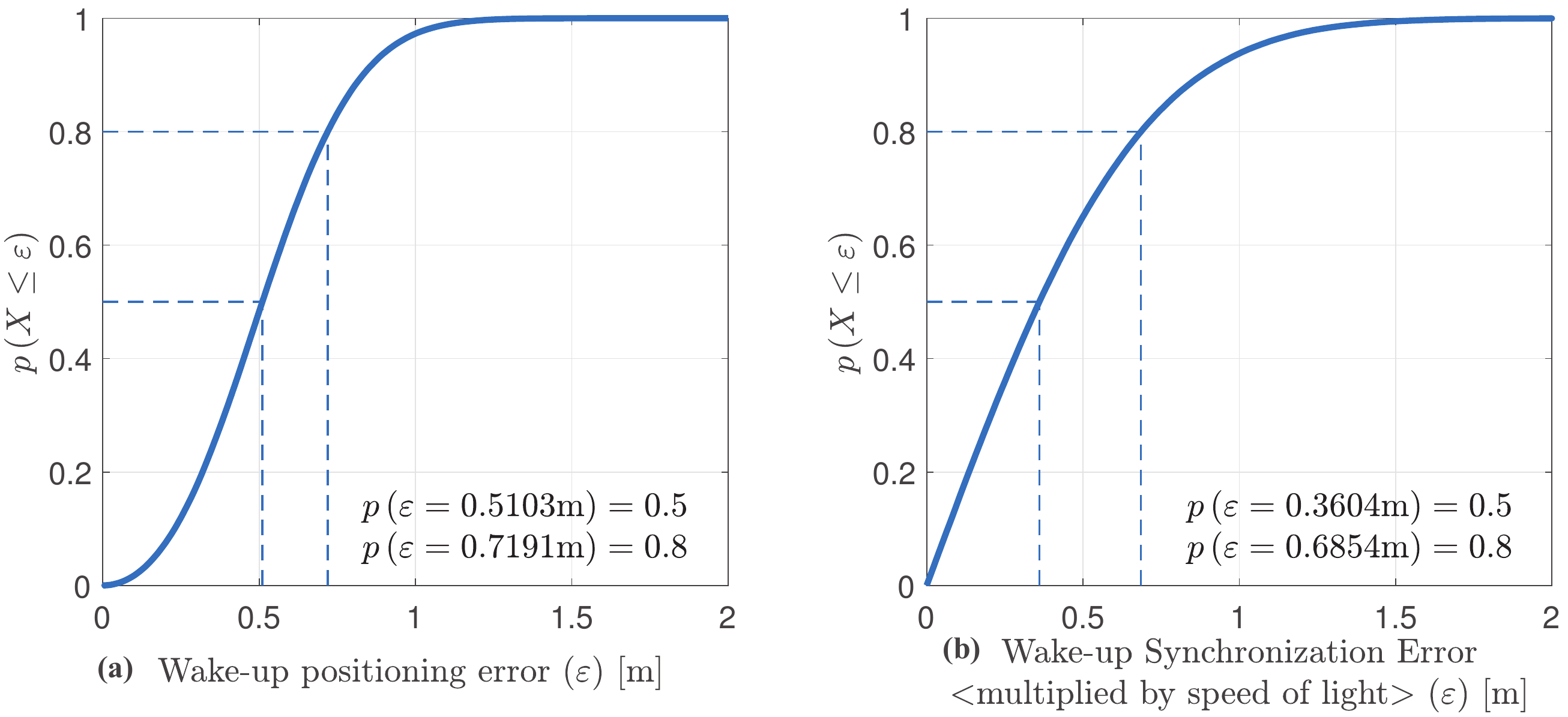}
  \caption{Performance of wake-up positioning and synchronization.
  }
  \label{fig_wakeup_performance}
  \vspace{-4ex}
\end{figure*}

\subsection{Radio Geometrization}
The radio geometrization component aims to characterize the exact geometric paths of the radio propagation rays from the base station to a newcomer vehicle, and further estimates the position and time synchronization bias of the newcomer vehicle. Fig. \ref{fig_scene} shows the estimation of radio geometrization of two newcomer vehicles from different directions, where the red lines indicate the real radio propagation paths, the blue lines denote the estimated radio propagation paths, the red crosses indicate the real position of the newcomer vehicles, and the green vehicles indicate the estimated position of the newcomer vehicles. Note that the time synchronization bias of a newcomer vehicle is also estimated in the radio geometrization component, which together with the position estimation of the newcomer vehicle is coined wake-up positioning and synchronization as described in Section \ref{sec_RG_SCWLS}.
The performance of wake-up positioning and synchronization can be seen in Fig. \ref{fig_wakeup_performance}, where the  wake-up positioning error can be seen in Fig. \ref{fig_wakeup_performance}(a) and the wake-up synchronization error can be seen in Fig. \ref{fig_wakeup_performance}(b). We can see that the 50 percentile error of wake-up positioning is $0.5103{\rm m}$, and the 50 percentile error of wake-up synchronization (multiplied by speed of light) is $0.3604{\rm m}$, which is much better than the initial positioning from GPS ($\sigma _G = 5 {\rm m}$). Fig. \ref{fig_scene} and Fig. \ref{fig_wakeup_performance} show that the radio geometrization component has a good performance in radio propagation path estimation as well as wake-up positioning and synchronization, where
the latter provides a more precise initial position and synchronization input to the cooperative SLAM component. That will improve the robustness of the TCSE algorithm when the GPS signals are in bad conditions or even dropped in urban scenario with bridges, high-rise buildings, or other obstructions.

\begin{figure*}
  \setlength{\abovecaptionskip}{-0.2cm}
  \centering
  \includegraphics[width=0.83\columnwidth]{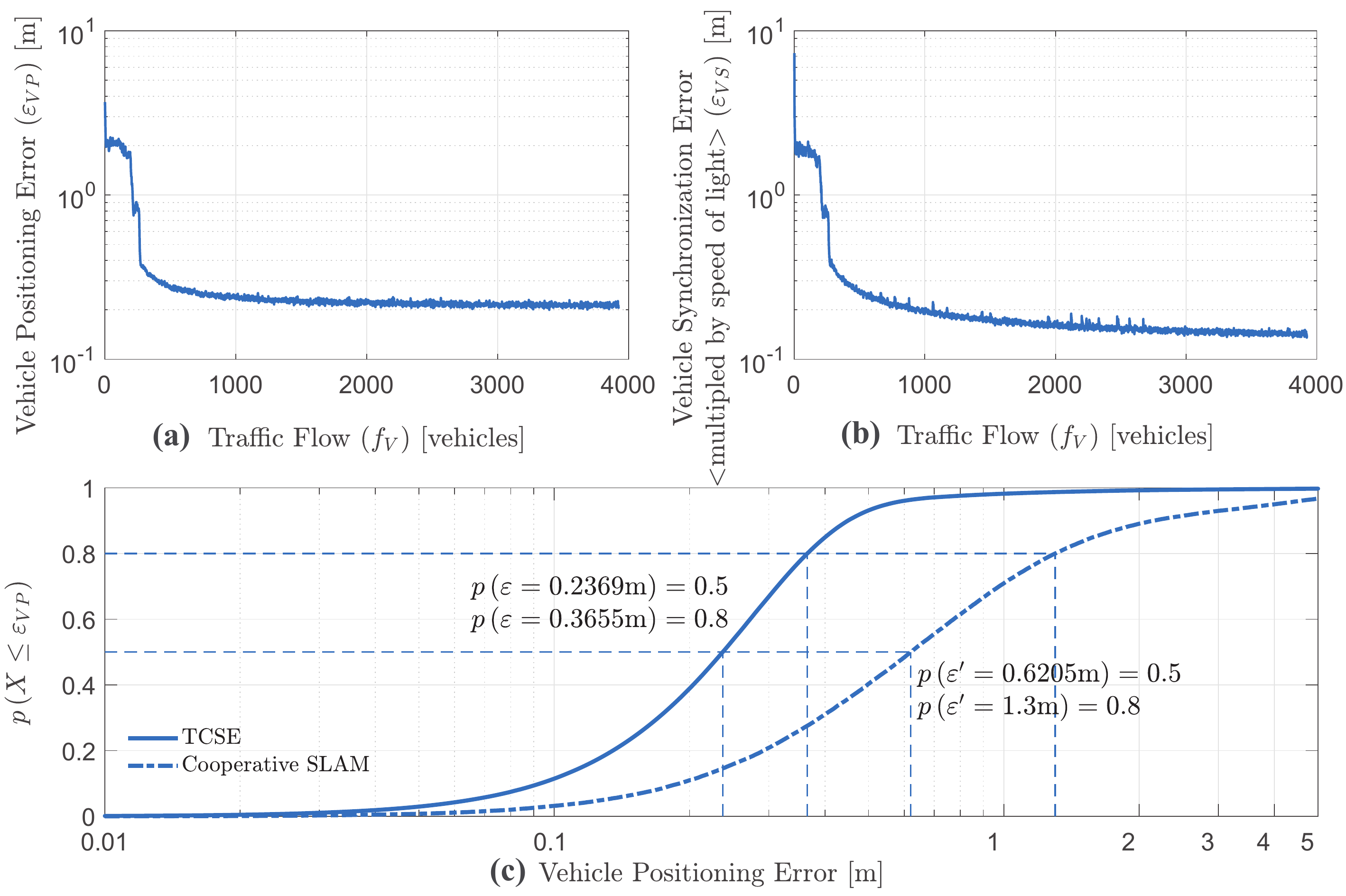}
  \caption{Performance of vehicle positioning and synchronization as the number of vehicles driving over time over the road section is increasing ((a) and (b)). The CDF of vehicle positioning error is shown in (c). As can be seen a dramatic gain of performance occurs when the cumulated number of vehicles over time exceeds 300, which allows a high-quality reconstruction of surrounding reflective structures.
  }
  \label{fig_vehicle_localization_performance}
  \vspace{-4ex}
\end{figure*}

\subsection{Vehicle Positioning and Synchronization}
\label{sec_SIM_4}
The simulation results of vehicle positioning and time synchronization can be seen in Fig. \ref{fig_vehicle_localization_performance}. For vehicle positioning, we can see from Fig. \ref{fig_vehicle_localization_performance}(a) and Fig. \ref{fig_vehicle_localization_performance}(c) that the vehicle positioning error gets gradually decreasing with the increasing of Vehicle Index and finally converges to a low error with 50 percentile value of $0.2369{\rm m}$. For vehicle synchronization, we can see from Fig. \ref{fig_vehicle_localization_performance}(b) that the vehicle synchronization error (multiplied by speed of light) also gets gradually decreasing with the increasing of Vehicle Index and finally converges to a low error with a mean value of  $0.1425{\rm m}$. {{What's more, compared with the situation without radio geometrization as well as reflective sensing and mapping ({\em Cooperative SLAM} in Fig. \ref{fig_vehicle_localization_performance}c), the TCSE has a much better performance in vehicle localization and synchronization.}}
This mainly results from: 1) the reliable data association method {{with a built-in ghost path elimination machine}} and joint estimation method for vehicles and CVTs in the cooperative SLAM component, 2) the sampling strategy in Section \ref{sec_CO_SLAM_IS} provides precise information of reflecting surfaces from the reflective sensing and mapping component to cooperative SLAM component, 3) the radio geometrization component provides precise initial position and time synchronization input to cooperative SLAM through wake-up positioning and synchronization. This further indicates that the TCSE can provide accurate vehicle positioning and time synchronization based on the interaction among the cooperative SLAM component, the reflective sensing and mapping component as well as the radio geometrization component.

\begin{figure*}
  \setlength{\abovecaptionskip}{-0.2cm}
  \centering
  \includegraphics[width=0.83\columnwidth]{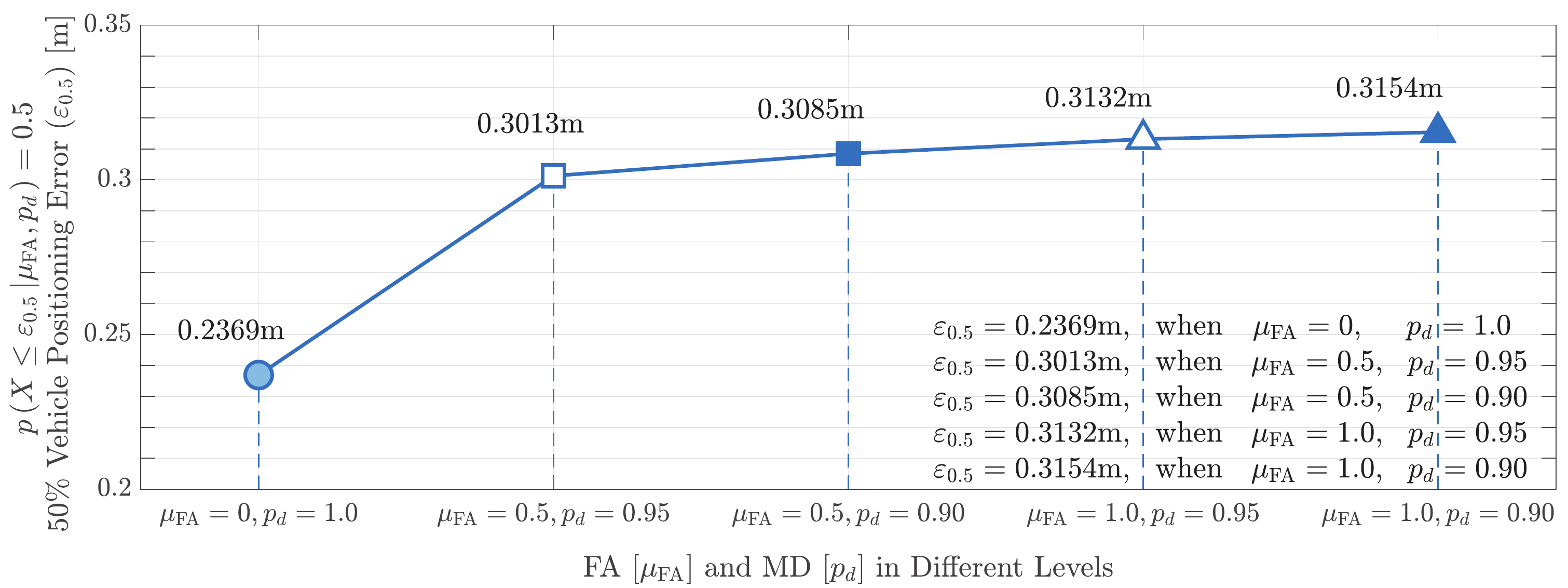}
  \caption{Performance of vehicle positioning over different levels of FA and MD.
  }
  \label{fig_vehicle_localization_performance_over_FA_MD}
  \vspace{-4ex}
\end{figure*}

  \subsection{Performance Against FA and MD}
  The performance of vehicle positioning over different levels of FA and MD is shown in Fig. \ref{fig_vehicle_localization_performance_over_FA_MD}. We can see from Fig. \ref{fig_vehicle_localization_performance_over_FA_MD} that the algorithm is fairly robust with respect to the FA and MD phenomenon, which mainly owns to the mechanism described in Section \ref{sec_CO_SLAM_CODA_3} allowing to check the consistency of the ghost path (falsely detected paths) with the rest of the data based on the reflective probability.

\section{Conclusion and Outlook}
\label{sec_conclusion}

We introduced the TCSE approach for joint vehicular localization and 3-D reflective mapping, which improves the accuracy of vehicular localization and its robustness to satellite-positioning signal conditions. Beyond improving the performance for localization, TCSE also has potential for communication performance enhancement. This is because the radio geometrization in TCSE would provide geometrical paths of radio propagation rays to beamforming designing for channel capacity improvement. An interesting open topic would for instance include the problem of joint communication and localization (sensing) performance enhancement following the example of recent works \cite{kakkavas2021power,luo2020multibeam,wymeersch2021integration}.

\vspace{-0.5cm}
\begin{appendices}
  \setcounter{equation}{0}
  \setcounter{subsection}{0}
  \renewcommand{\theequation}{A.\arabic{equation}}
  \renewcommand{\thesubsection}{A.\arabic{subsection}}
\section{Rotation Matrix}
\label{sec_appendix}

If rotating the vector ${{\bm r}_x} = {\left( {x,y,z} \right)^T}$ around the axis ${\bm e} = {\left( {{e_x},{e_y},{e_z}} \right)^T}$ by angle $\theta$, the rotated vector ${{\widehat {\bm r}}_x}$ can then be calculated as $  {{\widehat {\bm r}}_x} = {\mathfrak T}\left( {\theta ,{\bm e}} \right) \cdot {{\bm r}_x}$,
where ${\mathfrak T}\left( {\theta ,{\bm e}} \right)$ is defined as ${\mathfrak {\bm T}}\left( {\theta ,{\bm e}} \right)  = \widehat {\bm A} + \cos \theta  \cdot \left( {{\bm I} - \widehat {\bm A}} \right) + \sin \theta  \cdot {{\bm A}^*}$ with $\widehat {\bm A} = {\bm e} \otimes {{\bm e}^T}$ and ${{\bm A}^*} = \left( {0, - {e_z},{e_y};{e_z},0, - {e_x}; - {e_y},{e_x},0} \right)$.
Specially, if ${\bm e}$ is a set of normal vectors defined as ${\bm e} = \left( {{{\bm e}_1},{{\bm e}_2},...,{{\bm e}_N}} \right)$, then $\mathfrak{\bm T}\left( {\theta ,{\bm e}} \right) = diag\left\{ {\mathfrak{\bm T}\left( {\theta ,{{\bm e}_1}} \right),...,\mathfrak{\bm T}\left( {\theta ,{{\bm e}_N}} \right)} \right\}$.
\vspace{-0.3cm}
\section{Factor Graph Derivation for CVT-observation Data Association}
\label{sec_CODA_appendix}
This section shows the derivation of the joint probability distribution function for the CVT-observation association value $\varepsilon _n^{\left( k \right)}$ and the observation-CVT association value $o_m^{\left( k \right)}$, which can be described by the factor graph in Fig. \ref{CODA}.
{{\tiny
  \begin{align}
  & p\left( {o_m^{\left( k \right)},\varepsilon _m^{\left( k \right)}} \right) = \int {p\left( {\overset{\lower0.5em\hbox{$\smash{\scriptscriptstyle\frown}$}}{{\bm x}} _m^{\left( k \right)},{\bm r}_{\mathcal{N}_m^{\left( {k - 1} \right)}}^{\left( {k - 1} \right)},{\bm r}_{\left\{ {m,{p_m}} \right\}}^{\left( k \right)},{\mathcal{\bm R}_l},z_{\left\{ {m,{p_m}} \right\}}^{\left( k \right)},\rho _{\left\{ n \right\}}^{\left( {k - 1} \right)},o_m^{\left( k \right)},\varepsilon _m^{\left( k \right)}} \right){\text{d}}\overset{\lower0.5em\hbox{$\smash{\scriptscriptstyle\frown}$}}{{\bm x}} _m^{\left( k \right)}{\text{d}}{\bm r}_{\mathcal{N}_m^{\left( {k - 1} \right)}}^{\left( {k - 1} \right)}{\text{d}}{\bm r}_{\left\{ {m,{p_m}} \right\}}^{\left( k \right)}{\text{d}}{\mathcal{\bm R}_l}{\text{d}}z_{\left( {m,{p_m}} \right)}^{\left( k \right)}{\text{d}}\rho _{\left\{ n \right\}}^{\left( {k - 1} \right)}} \nonumber   \\
  & = \;\iint {\underbrace {\left\{ {\int {\prod\limits_{{p_m} = 1}^{{P_m}} {{h_{\left( {m,{p_m}} \right)}}\left( {o_{\left( {m,{p_m}} \right)}^{\left( k \right)} = 0\left| {\overset{\lower0.5em\hbox{$\smash{\scriptscriptstyle\frown}$}}{{\bm x}} _m^{\left( k \right)},{\bm r}_{\left( {m,{p_m}} \right)}^{\left( k \right)};z_{\left( {m,{p_m}} \right)}^{\left( k \right)},\mathcal{\bm R}} \right.} \right)} p\left( {{\bm r}_{\left\{ {m,{p_m}} \right\}}^{\left( k \right)}} \right)\left[ {\int_{l \in \mathcal{L}_{{\text{new}}}^{\left( {k - 1} \right)}} {{p_{\mathbbm 1}}\left( {{{\mathbbm 1}_{{\text{new}}}}\left| {\overset{\lower0.5em\hbox{$\smash{\scriptscriptstyle\frown}$}}{{\bm x}} _m^{\left( k \right)}} \right.;{\mathcal{\bm R}_l}} \right)p\left( {{\mathcal{\bm R}_l}} \right){\text{d}}{\mathcal{\bm R}_l}} } \right]{\text{d}}{\bm r}_{\left\{ {m,{p_m}} \right\}}^{\left( k \right)}} } \right\}}_{{\text{PDF of observation - CVT association value}}}} \nonumber  \\
  & \underbrace {\left\{ {\int {\prod\limits_{n \in \mathcal{N}_m^{\left( {k - 1} \right)}} {{g_{m,n}}\left( {\varepsilon _{m,n}^{\left( k \right)} = \left( {m,{p_m}} \right)\left| {\overset{\lower0.5em\hbox{$\smash{\scriptscriptstyle\frown}$}}{\bm x} _m^{\left( k \right)},{\bm r}_n^{\left( {k - 1} \right)}} \right.;z_m^{\left( k \right)},\mathcal{\bm R}} \right)} p\left( {r_{\mathcal{N}_m^{\left( {k - 1} \right)}}^{\left( {k - 1} \right)}} \right)\left[ {\int_{l \in \left\{ l \right\}} {{p_{\mathbbm 1}}\left( {{{\mathbbm 1}_{{\text{observe}}}}\left| {\overset{\lower0.5em\hbox{$\smash{\scriptscriptstyle\frown}$}}{{\bm x}} _m^{\left( k \right)}} \right.;{\mathcal{\bm R}_l}} \right)p\left( {{\mathcal{\bm R}_l}} \right)p\left( {z_{\left( {m,{p_m}} \right)}^{\left( k \right)}} \right){\text{d}}{\mathcal{\bm R}_l}} } \right] \cdot } } \right.}_{{\text{PDF of CVT - observation association value}}} \nonumber  \\
  & \underbrace {\left. { \cdot \left[ {\int_{l = \rho _n^{\left( {k - 1} \right)}} {\int {{p_\mathbb{O}}\left( {\mathbb{O}\left| {\overset{\lower0.5em\hbox{$\smash{\scriptscriptstyle\frown}$}}{\bm x} _m^{\left( k \right)};{\mathcal{\bm R}_l}} \right.} \right)p\left( {{\mathcal{\bm R}_l}} \right){\text{d}}{\mathcal{\bm R}_l}{\text{d}}\rho _n^{\left( {k - 1} \right)}} } } \right]{\text{d}}z_{\left( {m,{p_m}} \right)}^{\left( k \right)}{\text{d}}r_{\mathcal{N}_m^{\left( {k - 1} \right)}}^{\left( {k - 1} \right)}} \right\}}_{{\text{PDF of CVT - observation association value}}}\underbrace {\left\{ {\prod\limits_{n \in N_m^{\left( {k - 1} \right)}} {\prod\limits_{{p_m} = 1}^{{P_m}} {{\psi _{n,\left( {m,{p_m}} \right)}}} } } \right\}}_{{\text{global consistency constraint}}}p\left( {\overset{\lower0.5em\hbox{$\smash{\scriptscriptstyle\frown}$}}{{\bm x}} _m^{\left( k \right)}} \right){\text{d}}\overset{\lower0.5em\hbox{$\smash{\scriptscriptstyle\frown}$}}{{\bm x}} _m^{\left( k \right)} \label{eq_CODA_all}
\end{align}}}
\vspace{-0.5cm}
\section{Factor Graph Derivation for Reflector-CVT Data Association}
\label{sec_RCDA_appendix}
This section shows the derivation of the joint probability distribution function for the CVT-reflector association value $\rho _n^{\left( k \right)}$ and the reflector-CVT association value $\gamma_l^{\left( k \right)}$, which can be described by the factor graph in Fig. \ref{RCDA}.
{{\tiny
  \begin{align}
  & p\left( {\gamma _{\left\{ l \right\}}^{\left( k \right)},\rho _{\left\{ n \right\}}^{\left( k \right)}} \right) = \int {p\left( {{\bm r}_{\left\{ n \right\}}^{\left( k \right)},{\mathcal{\bm R}_{\left\{ l \right\}}},{\bm x}_{\mathcal{M}_n^{\left( k \right)}}^{\left( k \right)},z_{\mathcal{P}_n^{\left( k \right)}}^{\left( k \right)},\gamma _{\left\{ l \right\}}^{\left( k \right)},\rho _{\left\{ n \right\}}^{\left( k \right)}} \right){\text{d}}{\bm r}_{\left\{ n \right\}}^{\left( k \right)}{\text{d}}{\mathcal{\bm R}_{\left\{ l \right\}}}{\text{d}}{\bm x}_{\mathcal{M}_n^{\left( k \right)}}^{\left( k \right)}{\text{d}}z_{\mathcal{P}_n^{\left( k \right)}}^{\left( k \right)}}  = \int { \underbrace { \left\{ {\int {\prod\limits_{l = 1}^L {{u_l}\left( {\gamma _l^{\left( k \right)} = n\left| {{\bm r}_{\left\{ n \right\}}^{\left( k \right)},{\mathcal{\bm R}_l}} \right.} \right)} } p\left( {{\bm r}_{\left\{ n \right\}}^{\left( k \right)}} \right){\text{d}}{\bm r}_{\left\{ n \right\}}^{\left( k \right)}} \right\}}_{{\text{PDF of CVT - reflector}}\;{\text{association}}\;{\text{value}}}}  \nonumber \\
  &  \cdot \underbrace {\left\{ {\iint_{\begin{subarray}{l}
    m \in \mathcal{M}_n^{\left( k \right)} \\
    \left( {m,{p_m}} \right) \in \mathcal{P}_n^{\left( k \right)}
  \end{subarray}}  {{p_\mathbb{O}}\left( {\mathbb{O}\left| {{\bm x}_m^{\left( k \right)};{\mathcal{\bm R}_l}} \right.} \right)p\left( {{\bm x}_m^{\left( k \right)},z_{\left( {m,{p_m}} \right)}^{\left( k \right)}} \right){\text{d}}z_{\left( {m,{p_m}} \right)}^{\left( k \right)}{\text{d}}{\bm x}_m^{\left( k \right)}}} \right\}}_{{\text{PDF of reflector - CVT}}\;{\text{association}}\;{\text{value}}}\cdot \underbrace {\left\{ {\prod\limits_{l = 1}^L {\prod\limits_{n = 1}^N {{\varphi _{l,n}}} } } \right\}}_{{\text{global consistency constraint}}} p\left( {{\mathcal{\bm R}_{\left\{ l \right\}}}} \right){\text{d}}{\mathcal{\bm R}_{\left\{ l \right\}}} \label{eq_RCDA_all}
\end{align}}}

\end{appendices}

\ifCLASSOPTIONcaptionsoff
  \newpage
\fi

\vspace{-0.3cm}

\footnotesize
\bibliographystyle{IEEEtran}
\bibliography{IEEEabrv,reference}

\end{document}